\documentclass[aps,prd,nofootinbib, twocolumn, amsmath,amssymb]{revtex4-2}

\usepackage{amsmath, amsfonts, amssymb}
\usepackage{mathtools}
\usepackage{mathrsfs}
\usepackage{yhmath}
\usepackage{amssymb}
\usepackage{tensor}
\usepackage{accents}
\usepackage{tipa}
\usepackage{tikz}
\usetikzlibrary{shapes.geometric}
\usepackage{hyperref}
\usepackage{svg}

\usepackage{comment}

\usepackage{appendix}

\numberwithin{equation}{section}

\usepackage{setspace}
\usepackage[framemethod=TikZ]{mdframed}

\usepackage[T1]{fontenc}
\usepackage[english]{babel}
\usepackage{enumerate}
\usepackage{braket}
\usepackage{stackengine}
\usepackage[english]{babel}
\usepackage{amsthm}
\theoremstyle{plain}

\theoremstyle{plain}

\theoremstyle{plain}

\theoremstyle{definition}

\newcommand{\be}{\begin{equation}}
\newcommand{\ee}{\end{equation}}

\newcommand{\pd}{\partial}

\def\nn{\nonumber}

\def\p{\prime}

\def\d{\dot}

\def\dd{{\rm d}}

\def\ep{\epsilon}

\def\la{\lambda}
\def\om{\omega}

\def\De{\Delta}

\def\mL{\mathcal{L}}

\definecolor{newgreen}{rgb}{0.0, 0.75, 0.0}
\definecolor{cadmiumgreen}{rgb}{0.0, 0.42, 0.24}

\usepackage[framemethod=TikZ]{mdframed}
\usepackage{xcolor}

\stackMath
\allowdisplaybreaks

\hypersetup{%
  colorlinks=true,
  linkcolor=purple,
  linkbordercolor={0 0 1}
}

\begin{document}

\title{Correction to Hawking radiation in non-singular gravitational collapse}
\author{Hassan Mehmood} \email{hassan.mehmood@unb.ca}
\affiliation{Department of Mathematics and Statistics, University of New Brunswick, Fredericton, Canada.}
\date{\today}

\begin{abstract}
Recent studies have shown that quantum gravity introduces important corrections to the process of spherically symmetric gravitational collapse expected from general relativity. In particular, instead of falling into a central singularity, the collapsing body undergoes a bounce and eventually exits its Schwarzschild radius, and this entire process of collapse and rebound can occur in a single asymptotic region. In this paper, particle creation during such non-singular gravitational collapse is studied. It is shown that the probability of spontaneous emission of particles differs from the well-known probability of Hawking radiation from classical gravitational collapse. It is argued that the different result implies a deviation from thermality. Some arguments are also adduced concerning how the Hawking process during non-singular dust collapse could potentially remove shell crossing singularities.  
\end{abstract}

\maketitle

\section{Introduction}
In 1975 \cite{Hawking:1975vcx}, Stephen Hawking discovered the effect that has since been associated with his name: a spherically symmetric configuration of matter undergoing gravitational collapse spontaneously emits radiation which, at late times, will have a thermal spectrum. %While Hawking's calculations, strictly speaking, were valid only for an isolated Schwarzschild black hole, the generalization to a collapsing body which eventually forms the black hole being accomplished through heuristic arguments, his analysis can be refined, as was first done by Fredenhagen and Haag \cite{Fredenhagen:1989kr}. 
The framework in which Hawking's conclusions can be established is that of quantum field theory on curved spacetime. More precisely, three critical assumptions are used in %crucial assumptions can be teased out \cite{Ashtekar:2025ptw} from 
his analysis (or from any of its numerous reincarnations in the literature): (1) gravity is classical, i.e. the background spacetime in which all calculations are set up is described by a metric on a Lorentzian manifold; (2) the radiation which is spontaneously emitted by the collapsing body corresponds to the excitations of a quantum field which permeates the background spacetime, and these excitations behave as test particles (i.e., no back-reaction); and (3) the process of gravitational collapse is such as is predicted by the general theory of relativity. 

The first two assumptions constitute the essence of quantum field theory on curved spacetime. Before Hawking, it was already known through the work of Parker \cite{Parker:1968mv} and Imamura \cite{Imamura:1960tzx} that curved spacetimes can induce spontaneous emission of quanta of a quantum field. Indeed, it was understood from the earliest days of quantum field theory that particle production by spontaneous emission can occur in quantum fields in a strong electric field -- the so-called Schwinger effect \cite{Sauter-1931, Heisenberg:1936nmg, schwinger-1951}. Thus, it is not surprising that a gravitational background, including that describing a collapsing body, should induce particle production in a quantum field. What turned out to be surprising was the fact that, at late times, the spectrum of emitted particles is thermal, contradicting Hawking's own initial expectation of there being an initial burst of radiation which will eventually die out \cite[Chapter~2]{Hawking:1994ss}. This fact owes its origin to the third assumption referred to above. According to general relativity, given generic initial data, a reasonable matter configuration undergoing gravitational collapse in spherical symmetry will eventually form a Schwarzschild black hole \cite{choptuik-1993, Christodoulou-99, Wald:1997wa}. With this assumption, one can show that the late-time limit of the spectrum of particles emitted by the collapsing matter is thermal \cite{Fredenhagen:1989kr}. (There are many complementary ways to understand the origin of thermality in radiation from a black hole. The explanation here is tailored for the purposes of this paper; for alternative perspectives, see, e.g., Refs. \cite{Witten:2024upt, Padmanabhan:2009vy}.) 

\begin{figure}[ht]

\tikzset{every picture/.style={line width=0.6pt}} %set default line width to 0.75pt        

\begin{tikzpicture}[x=0.6pt,y=0.6pt,yscale=-1,xscale=1]
%uncomment if require: \path (0,589); %set diagram left start at 0, and has height of 589

%Shape: Triangle [id:dp10862430376989674] 
\draw   (545,288.04) -- (359,473) -- (359,101.5) -- cycle ;
%Curve Lines [id:da75983791644928] 
\draw [color={rgb, 255:red, 74; green, 144; blue, 226 }  ,draw opacity=1 ][line width=1.5]    (359,101.5) .. controls (369.71,127.85) and (462.62,256.15) .. (450.81,266.07) .. controls (439,276) and (349,279) .. (374,302) .. controls (399,325) and (395,325) .. (399,336) .. controls (403,347) and (372,445) .. (359,473) ;
%Straight Lines [id:da9312958391283306] 
\draw [color={rgb, 255:red, 208; green, 2; blue, 27 }  ,draw opacity=1 ]   (450.81,266.07) -- (393,323) ;
%Curve Lines [id:da11718406569976048] 
\draw [color={rgb, 255:red, 208; green, 2; blue, 27 }  ,draw opacity=1 ]   (393,323) .. controls (395,325) and (371,311) .. (369,295) .. controls (367,279) and (411,273) .. (450.81,266.07) ;

% Text Node
\draw (352,475.4) node [anchor=north west][inner sep=0.75pt]    {$i^{-}$};
% Text Node
\draw (353,81.4) node [anchor=north west][inner sep=0.75pt]    {$i^{+}$};
% Text Node
\draw (551,277.4) node [anchor=north west][inner sep=0.75pt]    {$i^{0}$};
% Text Node
\draw (471,364.4) node [anchor=north west][inner sep=0.75pt]    {$\mathcal{J}^{-}$};
% Text Node
\draw (452,174.4) node [anchor=north west][inner sep=0.75pt]    {$\mathcal{J}^{+}$};
% Text Node
\draw (307,272.4) node [anchor=north west][inner sep=0.75pt]    {$r=0$};
\draw (420,292.4) node [anchor=north west][inner sep=0.75pt]    {$\textcolor[rgb]{0.82,0.01,0.11}{H_{O}}$};
% Text Node
\draw (390,255.4) node [anchor=north west][inner sep=0.75pt]    {$\textcolor[rgb]{0.82,0.01,0.11}{H_{I}}$};

\end{tikzpicture}
    \caption{Example of non-singular gravitational collapse. The blue line represents the boundary of the collapsing star and the region inside the red figure is trapped.}
    \label{fig:non-sing-collapse-1}
\end{figure}
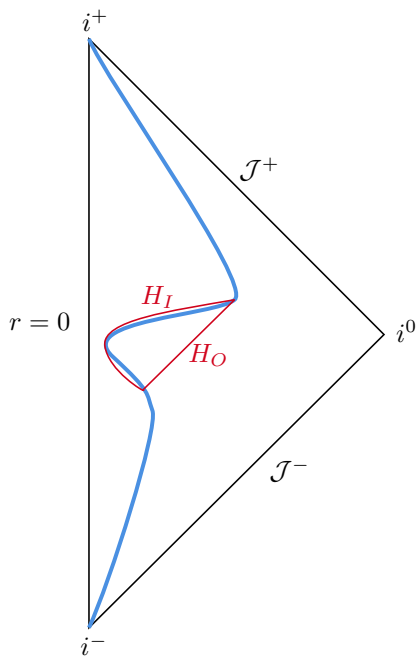

In view of the foregoing discussion, one may ask the following question: to what extent does the spectrum of radiation change if the picture of classical gravitational outlined above is modified? In this connection one must first ask as to why one might expect the classical picture to change. Such an expectation arises from considerations concerning quantum gravity. One of the aims of formulating a satisfactory theory of quantum gravity is to understand the behavior of nature in domains where classical theories of gravity fail. One such domain is the very late stage of gravitational collapse, for then the matter inside the (trapping) horizon approaches the central singularity where general relativity breaks down. Just as the development of a fully consistent quantum theory of electrodynamics sounded a death knell for the infinities plaguing the classical theory of electrodynamics, one expects a fully consistent theory of quantum of gravity to cure the singularities of classical general relativity, and thereby change the conventional picture of such processes as gravitational collapse and the big bang. In fact, investigations into these processes from a variety of quantum theoretical perspectives have yielded promising results. For instance, quantum mechanical evolution of a collapsing thin shell of matter \cite{Saini:2014qpa, Greenwood:2008ht, Wang:2009ay, Torres:2014pea} leads to the resolution of the central singularity. More complicated matter profiles require a field-theoretical analysis, which can be accomplished within the framework of various candidate theories of quantum gravity, such as asymptotically free quantum gravity \cite{Bambi:2013gva, Bambi:2016uda, Liu:2014kra} and loop quantum gravity (LQG) \cite[and references therein]{Husain:2021ojz, Husain2024, Wilson-Ewing:2016yan, Cipriani:2024nhx, Bojowald:2024ium, Ashtekar:2011ni, Haggard:2014rza, Han:2023wxg, Giesel:2023hys, Giesel:2023tsj, Alonso-Bardaji:2025hda}. The common lesson which emerges from these studies is that gravity becomes repulsive at the Planck scale, portending the demise of cosmological and black-hole singularities, replacing them instead with a bounce in the geometry at a critical radius. Of particular interest are the detailed studies of quantum gravitational collapse from the perspective of asymptotic freedom \cite{Bambi:2013gva, Bambi:2016uda, Liu:2014kra} and of LQG \cite{Husain:2021ojz, Husain:2022gwp, Cipriani:2024nhx}. Despite the difference in approach, both sets of investigation lead to a similar picture of quantum-corrected gravitational collapse (Fig.~\ref{fig:non-sing-collapse-1}): matter undergoing collapse falls into a transient trapping horizon, thus forming a black hole, bounces outward upon reaching a minimum radius proportional to the Planck length, and eventually exits the trapping horizon, thus bringing the black hole's life to an end \footnote{Refs. \cite{Husain:2021ojz, Husain:2022gwp, Cipriani:2024nhx} also predict a shockwave propagating as a discontinuity in the spacetime geometry. We shall ignore this complication in our computations; however, in the concluding section, we speculate on the potential implications of incorporating the shock-wave discontinuity into the analysis of black hole evaporation.} -- all of this occurring in a single asymptotic region. This picture of gravitational collapse is radically different from the classical one, and it is the aim of this paper to study how the former affects the spectrum of radiation from the collapsing body.   

%The understanding that gravitational collapse should lead to the formation of a black hole characterized by the presence of a (trapping) horizon behind which lies a singularity pervades the literature on black-hole thermodynamics. In so far as one is interested in exploring the relationship between a classical description of gravitational collapse and thermodynamics, this understanding is justified. However, if the aim is to understand the implications of a quantum theory of gravity for 

%Inspired by recent work on quantum-corrected gravitational collapse \cite{Husain:2022gwp}, we wish to study some thermodynamical aspects of the kind of non-singular gravitational collapse depicted in Fig~\ref{fig:non-sing-collapse-1}. That is, we shall look at the spectrum of particles produced by the black region bounded by the red figure, study the (time-dependent) temperature and entropy of the black hole and the radiation, and seek to generalize, if possible, the thermodynamical laws that hold for classical stationary black holes in the presence of a quantum field. 

The analysis of radiation in a general dynamical spacetime is a nontrivial task. In classical gravitational collapse, as alluded to above, a tractable result obtains in the form of a thermal spectrum because of the late-time behavior of the spacetime geometry. In the context of a collapse scenario of the kind depicted in Fig.~\ref{fig:non-sing-collapse-1}, the late-time limit of the spacetime has no black hole at all; instead, the black-hole is characterized by a trapping horizon which exists for only a finite amount of time. Thus, such tools are required as enable one to study radiation from arbitrary trapping horizons, which are local and dynamical \cite{hayward-bh-dynamics, hayward-bh-spherical, Ashtekar-Krishnan-2003, Ashtekar:2025-horizons}, unlike the event horizon of a Schwarzschild black hole, which is static and non-local. These tools are most easily developed in a framework which treats Hawking radiation as the separation by the gravitational background of virtual particle-antiparticle pairs into real particles and antiparticles \cite{Vanzo:2011wq}. We will use this framework to calculate, in the semiclassical approximation, the probability of spontaneous emission by the trapped region (black hole) in the kind of non-singular gravitational collapse in Fig.~\ref{fig:non-sing-collapse-1}. The main result \eqref{eq:prob-emission} is that now both the inner ($H_I$) and outer ($H_O$) parts of the trapping horizon potentially contribute to the emission probability. This suggests that, unlike the classical case, the spectrum of radiation from a non-singular gravitational collapse may not be thermal. Furthermore, the modus operandi of pair creation by black holes, with the particle escaping to future infinity and the antiparticle being absorbed by the black hole, suggests a possible mechanism by which shell crossing singularities in the gravitational collapse of dust might be removed by the Hawking process. Such a mechanism is outlined in the concluding section.  

The paper is organized as follows. In Section~\ref{sec:bh-dynamics}, we review the class of metrics which describe and the tools which enable us to study dynamical black holes, defined as the regions bounded by future trapping horizons -- the discussion will be restricted to spherical symmetry. In Section~\ref{sec:qm-rel-particle}, the proper-time formalism of Feynman \cite[Appendix~A]{Feynman:1950ir} that recasts scalar quantum field theory in terms of the quantum mechanics of a relativistic particle is reviewed, since it forms a precursor to understanding the use of tunneling methods in studying Hawking radiation. These methods are reviewed in Section~\ref{sec:rad-tunneling-methods}, and used to calculate the probability of spontaneous emission of particles by the trapped region in Fig.~\ref{fig:non-sing-collapse-1}. Finally, some conclusions from our study are noted in the concluding section. Unless otherwise stated, we work in units in which $G=c=1$.

\section{Black-hole dynamics}\label{sec:bh-dynamics}
One of the simplest settings to study gravitational collapse in spherical symmetry is that of (Laimetre-Tolman-Bondi) LTB spacetimes \cite{Bondi:1947fta, Tolman:1934za, Lemaitre:1931zza}, which describe a universe filled with dust (pressureless fluid). A quantum-corrected version of LTB collapse using LQG methods can be most conveniently studied in generalized Painleve-Gullstrand (PG) coordinates \cite{Husain:2021ojz, Husain:2022gwp, Cipriani:2024nhx}. The marginally bound case is described by the metric  
\begin{equation}
    \dd s^2 = -\dd t^2 + (\dd r+ N^r(t,r) \dd t)^2 + r^2\dd\Omega^2~. \label{eq:pg-metric}
\end{equation}
where $N^r$, the radial shift, further depends on a ``mass function'' $M(t,r)$, which approaches the conserved ADM mass $M_{\text{ADM}}$ as $r\to\infty$,
\begin{equation}
    N^r = \sqrt{\frac{2M}{r}}
\end{equation}
The precise form of $M$ will determine the precise details of how matter behaves (the blue line in the Fig.~\ref{fig:non-sing-collapse-1}). Examples of mass functions with the bouncing behavior shown in the figure can be found in Refs. \cite{Hergott:2022hjm, Husain:2022gwp, Husain:2021ojz}. Our calculations will only depend on the generic bouncing behavior of $M(t,r)$, regardless of its precise form. Nevertheless, to spell out the kind of assumptions involved in obtaining a quantum-corrected bouncing model, let us consider a specific example, namely quantum-corrected, marginally bound dust collapse in the framework of effective LQG \cite{Husain:2022gwp}.

\subsection{Quantum-corrected dust collapse in LQG}
In canonical LQG, the classical starting point is a Hamiltonian formulation of general relativity in terms of $\mathfrak{su}(2)$-valued triads and the Ashtekar-Barbero connection, defined on the spatial hypersurfaces of an arbitrary $3+1$ foilation of a four-dimensional spacetime \cite{Ashtekar:2004eh}. One then considers the integrals of the triads over two-dimensional surfaces and holonomies of the Ashtekar-Barbero connection around edges embedded in the spatial hypersurfaces. These objects satisfy the so-called holonomy-flux algebra, which is then represented on a Hilbert space possessing a background-independent inner product. The solution to the quantum theory then consists in representing the diffeomorphism and Hamiltonian constraints of general relativity as operators on the background-independent Hilbert space and à la Dirac, finding the null space of these operators\footnote{This part of the LQG program is still not fully complete, the fundamental problem being a fully satisfactory quantization of the Hamiltonian constraint (see Ref.~\cite{Ashtekar:2017yom} for more details).}. The semiclassical limit of the theory can be obtained by finding the expectation values of the holonomy, flux and constraint operators and the quantum equations of motion in suitably defined coherent states \cite{Thiemann:2000bw}. In principle, this should give rise to an effective Hamiltonian field theory which incorporates corrections of order $\hbar$ or higher in the effective equations of motion. In practice, however, taking the semiclassical limit in this way is a delicate business, for one has to deal carefully with problems concerning covariance, anomalies and gauge-fixing (see Ref.~\cite{Bojowald:2012xy} for an extensive review in the case of quantum cosmology). However, in the case of marginally bound dust collapse with the gauge-fixed metric \eqref{eq:pg-metric}, there is a consistent effective theory, which we now briefly describe\footnote{For more general frameworks within LQG, see Refs.~\cite{Giesel:2023tsj,Zhang:2024khj, Alonso-Bardaji:2025hda}.} \cite{Husain:2022gwp}.

In marginally bound dust collapse, the Einstein equations in PG coordinates reduce to a single equation for the mass function $M(t,r)$, namely
\begin{equation}
    \pd_t M_c -\sqrt{\frac{2M_c}{r}}\pd_r M_c = 0~,
\end{equation}
where $c$ indicates that we are referring to a classical function (as opposed to a quantum-corrected function in the effective theory). In this gauge-fixed setting, the triad variables of canonical LQG are all gauge-fixed, whereas the Ashtekar-Barbero connection reduces to a single function $b(t,r)$, the component of the connection which gives rise to holonomies in the $\theta$ direction. It is related to the mass function\footnote{For simplicity, we set the Barbero-Immirzi parameter to 1.} via
\begin{equation}
    M_c(t,r) = \frac{rb^2}{2}
\end{equation}
and hence the equation of motion in terms of $b$ is\footnote{In writing down the effective equation of motion for $b$, we have chosen the positive root of $M=rb^2/2$. Thus, the $b$ variable that we use differs from that in Ref.~\cite{Husain:2022gwp} by a minus sign.}
\begin{equation}
    \pd_t b - \frac{1}{2r}\pd_r(rb^2) = 0~.
\end{equation}
In passing to the quantum-corrected effective theory, this equation is replaced by \cite{Husain:2022gwp}
\begin{equation}
    \pd_tb - \frac{1}{2r\mu}\pd_r\left( r^3 \sin^2\frac{\sqrt{\mu}b}{r} \right) = 0~, \label{eq:eom-effective-lqg}
\end{equation}
where $\mu$ is a constant proportional to the square of the Planck length (i.e., $\hbar$ in $G=c=1$ units). From this, we can extract an effective mass function
\begin{equation}
    M(t,r) = \frac{1}{2\mu}r^3\sin^2\left(\frac{\sqrt{\mu}b}{r}\right)~. \label{eq:mass-in-b-eff}
\end{equation}
It can be verified that in the limit $\mu \to 0$, this function reduces to the classical mass function $M_c$.
It is also clear how this $\mu$-dependent quantum correction can tame classical singularity. From the fact that $\sin^2(x)\leq 1$ for all $x\in\mathbb{R}$, it follows that
\begin{equation}
    M \leq \frac{r^3}{2\mu} \label{eq:mass-bound}
\end{equation}
Thus, the mass function is bounded from above, unlike the classical case. The same conclusion can be reached with regard to the energy density of dust, which in PG coordinates is given by
\begin{equation}
    \rho(t,r) = \frac{\pd_r M}{4\pi r^2} = \frac{1}{8\pi\mu r^2}\pd_r\left(r^3\sin^2\frac{\sqrt{\mu}b}{r}\right)~.
\end{equation}
This also allows us to write \eqref{eq:eom-effective-lqg} as
\begin{equation}
    \pd_t b = 4\pi r\rho~,
\end{equation}
which is useful because it establishes that $b$ is a monotonically increasing function of time (since we may assume $\rho\geq 0$ for all physical matter profiles). We can also extract the domain of $b$ by taking the square-root of \eqref{eq:mass-in-b-eff},
\begin{equation}
    \sin(\sqrt{\mu}b/r) = \sqrt{2\mu M/r^3}~,
\end{equation}
the positive root chosen in agreement with \eqref{eq:eom-effective-lqg}. Since $M\geq 0$ for physical matter, such as dust,
\begin{equation}
    \frac{\sqrt{\mu}b}{r} \in [0,\pi]~. 
\end{equation}

To understand the qualitative features of the model, it is instructive to look at a specific mass profile, such as that for the Oppenheimer-Snyder model of stellar collapse \cite{Husain:2022gwp}. In that case, there is a moving boundary $r=L(t)$ of the star inside which the energy density of dust is radially constant and outside which it is zero. Thus, for $r> L(t)$, we have
\begin{equation}
    \rho = 0 \Rightarrow M_e(t,r) = M_{\text{ADM}}~,
\end{equation}
where the $e$ in the subscript refers to the exterior region ($r>L(t)$). This implies that in the exterior of the star, $b$ is given by the equation
\begin{equation}
    \sin\left(\frac{\sqrt{\mu}b_e}{r}\right) = \sqrt{\frac{2\mu M_{\text{ADM}}}{r^3}}~,
\end{equation}
Since the right-hand side is a positive constant in time, to make sure that $b_e \in [0,\pi]$, we can have either
\begin{equation}
    b_e = \frac{r}{\sqrt{\mu}}\arcsin\left(\sqrt{\frac{2\mu M_{\text{ADM}}}{r^3}}\right)
\end{equation}
or the same expression subtracted from $\pi$. We choose the former. In the interior region $(r < L(t))$, $\pd_r\rho = 0$, which implies that $\pd_r (\sqrt{\mu}b/r) = 0$. Thus, \eqref{eq:eom-effective-lqg} becomes \cite{Husain:2022gwp}
\begin{equation}
    \pd_t \left(\frac{\sqrt{\mu}b_i}{r}\right) = \frac{3\sin^2(\sqrt{\mu}b_i/r)}{2\sqrt{\mu}}~,
\end{equation}
where the $i$ in subscript refers to the interior of the star. This equation has the solution
\begin{equation}
    \sin(\sqrt{\mu}b_i/r) = \frac{1}{\sqrt{1+9(t-t_0)^2/4\mu}}~ \label{eq:internal-os-sol-b}
\end{equation}
for some constant $t_0$. To determine $b_i$ from this equation, some care is required. The right-hand side of the preceding equation is now time-dependent, and in particular, increases in time for $t<t_0$ and decreases in time for $t>t_0$. Therefore, since $b$ must always be monotonically increasing in time, $b_i \in [0,\pi/2]$ for $t<t_0$ and $b_i\in [\pi/2,\pi]$ for $t>t_0$. That is, we must have 
\begin{equation}
    b_i = \begin{cases}
        \arcsin\left(\frac{1}{\sqrt{1+9(t-t_0)^2/4\mu}}\right)~, & t < t_0\\
        \pi - \arcsin\left(\frac{1}{\sqrt{1+9(t-t_0)^2/4\mu}}\right)~, & t > t_0
    \end{cases}
\end{equation}
To determine the evolution of the boundary $L(t)$ of the star, one can try to match the external and internal solution. Doing this for $t<t_0$ yields
\begin{equation}
    L(t)^3 = \frac{M_{\text{ADM}}}{2}\left(4\mu + 9(t-t_0)^2\right) \label{eq:L-internal}
\end{equation}
However, this only holds $t<t_0$, for then both $b_i, b_e\in[0,\pi/2]$, whereas for $t>t_0$, $b_e\in[0,\pi/2]$ while $b_i\in[\pi/2,\pi]$, making it impossible to impose continuity on $b$. This is a manifestation of the fact that \eqref{eq:eom-effective-lqg} is a hyperbolic conservation law in the variable $B=rb$, and smooth solutions to hyperbolic conservation laws can develop discontinuities (called shock waves in the technical literature) in a finite time \cite{leveque}. To the find the solution beyond the appearance of the first discontinuity, one has to integrate the conservation law over $[S-\ep, S+\ep]$, where $r=S$ is the location of the shock wave. This leads to an evolution equation for the shock wave, the so-called Rankine-Hugoniot (RH) condition
\begin{equation}
    \frac{\dd S}{\dd t} = \frac{S(t)^2}{2\mu}\frac{M(t,b^+)-M(t,b^-)}{b^+-b^-}~, \label{eq:rh-condition}
\end{equation}
where $b^{\pm}=\lim_{r\to S^{\pm}}b$. Although this equation is necessary to find the correct dynamics in the rebound phase, for simplicity, whenever using the Oppenheimer-Snyder model, we will assume $L(t)$ to have the same form in the post-bounce phase ($t>t_0$) as it has in the pre-bounce phase ($t<t_0$), i.e., eqn.~\eqref{eq:L-internal}. This simplifying assumption will change the black-hole life time \cite{Husain:2022gwp} from what it should turn out to be under \eqref{eq:rh-condition}, but this equation is just one possible instance of the RH condition. In general, there are many inequivalent ways of integrating a conservation law, and each leads to a different RH condition, which in turn leads to a different lifetime for the black hole \cite{Fazzini:2025hsf}. In the concluding section, we shall outline an argument suggesting that Hawking radiation from the collapsing body might disperse the discontinuity in the mass function and hence one may be able to dispense with the need of using the RH condition altogether.

%and in the current example, this equation is found to be \cite{Husain:2022gwp}
%\begin{equation}
 %   L(t) = \left( \frac{3\sqrt{\mu}M_{\text{ADM}}(t-t_0)}{\pi} \right)^{1/3} + \mO(L_\text{min}/L)~,
%\end{equation}
%where $L_{\text{min}} = (2\mu M_{\text{ADM}})^{1/3}$, and the equation is valid for $t>t_0$.}

Notice that \eqref{eq:L-internal} entails that the falling matter does not fall to $r=0$, as in the classical case, but reaches the minimum radius $L_{\text{min}}$, and then rebounds and starts falling outwards. This is also clear from the behavior of the mass function and energy density inside the star. From \eqref{eq:internal-os-sol-b}, we find that
\begin{equation}
    M_i(t,r) = \frac{2r^3}{4\mu + 9(t-t_0)^2}~ \label{eq:mass-inner-os}
\end{equation}
and hence
\begin{equation}
    \rho_i = \frac{2}{2\pi(4\mu+9(t-t_0)^2)}~.
\end{equation}
The bouncing nature of the geometry is manifest. For fixed $r$, say $r=r^*$, as $t\to t_0^+$, $M(t,r)$ increases, i.e., more and more mass piles up in the region $r \leq r^*$. When $t=t_0$, the mass function attains its maximum value $r^{*3}/2\mu$. And for $t>t_0$, the mass function decreases with time, indicating that the dust accumulated behind $r^*$ has now started to move out, i.e. the star has bounced. It is also worth emphasizing that the bounce is of quantum origin: it is the constant $\mu$, proportional to $\hbar$, which ensures that the energy density in the interior undergoes a bounce at $t=t_0$ rather than diverging as $t\to t_0^+$, as in the classical theory\footnote{In Ref.~\cite{Husain:2022gwp}, which we have closely followed in this subsection, the shift function $N^r$ is different in the effective theory from what we have assumed. We have introduced a quantum correction in the equation of motion for $b$, which modifies the mass function $M$, which in turn modifies the shift function through the classical relation $N^r=\sqrt{2M/r}$. In Ref.~\cite{Husain:2022gwp}, however, the quantum correction to $N^r$ is introduced independently of the correction to $b$, and this renders the relation $N^r=\sqrt{2M/r}$ in their framework. We have chosen to avoid this path because (1) the effective theory allows it (i.e., it is consistent) and (2) the relation $N^r=\sqrt{2M/r}$ leads to a very simple equation for the location of the horizon, as shown in the next subsection.}.

\subsection{Dynamical horizons}
Let us now return to a generic mass function $M(t,r)$, and ask how such features as the location of trapped regions and the black-hole boundary can be studied in the context of a general time-dependent spacetime that describes gravitational collapse. To this end, we shall follow the ideas first laid out in Refs.~\cite{hayward-bh-dynamics, hayward-bh-spherical}. Let us begin by observing that due to spherical symmetry, every $(t,r)$ corresponds to a 2-sphere of symmetry, and every such sphere has two independent -- one ingoing and one outgoing -- null vectors orthogonal to the sphere. To write down these null vectors, note that along a radial null ray
\begin{align*}
    \mathrm{d} s^2 = 0 &= -\mathrm{d} t^2 + (\mathrm{d} r+N^r\mathrm{d} t)^2  \Rightarrow  \frac{\mathrm{d} r}{\mathrm{d} t} = \pm 1 - N^r~,
\end{align*}
where the plus (minus) sign refers to outgoing (ingoing) rays. This invites us to define
\begin{equation}
    n^a_{+} = \frac{1}{\sqrt{2}}(1, 1-N^r, 0, 0) ~, \quad n^a_{-} = \frac{1}{\sqrt{2}}(1, -1-N^r, 0, 0)~.
\end{equation}
These are manifestly orthogonal to the 2-metric $r^2\dd\Omega^2$; the prefactor of $1/\sqrt{2}$ ensures $n^a_{+} n_{-a} = -1$, for convenience. Using these, we can find the outgoing and ingoing null expansions, which are given by
\begin{align*}
    \theta_{\pm} = q_{ab}\nabla^an^b_{\pm} ~,
\end{align*}
where $q_{ab} = g_{ab} + \tfrac{1}{2}n_{+(a}n_{-b)}$ is the induced metric on the 2-spheres of symmetry. In our chosen coordinates, we get
\begin{equation}
    \theta_{\pm} = \pm(1 \mp N^r ) \label{eq:theta-pm}
\end{equation}
The trapped region shown in Fig.~\ref{fig:non-sing-collapse-1} is bounded by the future trapping horizon, which is defined as the region where the expansion of outgoing null geodesics vanishes, $\theta_+ = 0$, and the expansion of ingoing null geodesics is negative, $\theta_{-} < 0$; these two conditions are met if $N^r=1$, which is true at the hypersurface $r=r_H(t)$ given by
\begin{equation}
    r_H(t) = 2M(t, r_H(t)) \equiv 2M_H~. 
\end{equation}
This implies that
\begin{equation}
    \frac{\dd r_H}{\dd t} = 2\d{M}_H + 2M^{\p}_H\frac{\dd r_H}{\dd t}~, \qquad \d{f}\equiv \frac{\partial f}{\partial t}~, \quad f^{\p} \equiv \frac{\partial f}{\partial r} ~.  
\end{equation}
Therefore,
\begin{equation}
    \frac{\dd r_H}{\dd t} = \frac{2\d{M}_H}{1-2M^{\p}_H}~.
\end{equation}
Thus, the metric \eqref{eq:pg-metric} at the horizon becomes
\begin{align}
    \dd s^2|_{H} &= \frac{\dd r_H}{\dd t}\left(\frac{\dd r_H}{\dd t} + 2 \right)\dd t^2~ + r_H\dd\Omega^2\\
    &= \frac{4\d{M}_H}{(1-2M^{\p}_H)^2}\left[1 + (\d{M}_H - 2M^{\p}_H)  \right]\dd t^2 + r_H\dd\Omega^2~. \label{eq:metric-horizon-1}
\end{align}
Outside the collapsing star, and also at its boundary $r=r_b(t)$, $M(t,r) = M_{\text{ADM}}$, and so $\d{M}_H = M^{\p}_H = 0$, which entails that $\dd s^2|_{H} = 0$ for constant angular coordinates; this explains why $H_O$, the outer part of trapping horizon, which is in the vacuum region, is shown to be null in Fig.~\ref{fig:non-sing-collapse-1}; in technical terms, it is a non-expanding horizon \cite{Ashtekar:2025ptw}. The rest of the trapping horizon will have both spacelike and timelike regions, depending on the mass function. Since $\d M_{H_I} < 0$ ($\d M_{H_I} > 0$) in the collapse (rebound) phase, we should generally expect the inner part $H_I$ of the horizon to be timelike before the bounce and spacelike after the bounce.

Eqn.~\eqref{eq:metric-horizon-1} can be written in a much more illuminating form. In spherical symmetry, there always exists a vector field $K^a$, called the Kodama vector field \cite{Kodama:1979vn, hayward-bh-spherical}, such that $\nabla^a(G_{ab}K^b) = 0$. In PG coordinates, it is given by
\begin{equation}
    K^a = (1, 0, 0, 0)~,
\end{equation}
as can be immediately verified. Its norm is given by
\begin{equation}
    K^aK_a = -1+ N^r(t,r)^2 = \theta_{+}\theta_{-}~.
\end{equation}
Inside the trapped the region, $\theta_{+}, \theta_{-} < 0$, and so $K^a$ is spacelike, while at the future trapping horizon, it is null. Thus, $K^a$ behaves similarly to the asymptotically timelike Killing vector field in a static spherically symmetric spacetime. It can be analogously used to define a notion of surface gravity in the more general case. Indeed, on the trapping horizon, one has
\begin{equation}
    K^a\nabla_{[b}K_{a]} = -\kappa_{H} K_b
\end{equation}
for some function $\kappa_H$ defined on the trapping horizon. In the static case, $K^a$ becomes the asymptotically timelike Killing field and hence $\nabla_{(a}K_{b)} =0$, so that the above equation reduces to the standard definition of surface gravity for a static black hole. Thus, we identify $\kappa_H$ as the surface gravity of the black hole. Using the preceding equation, one can show that
\begin{equation}
    \kappa_{H} = \frac{1}{2} g^{AB}\nabla_{A}\nabla_{B}r~|_{H},
\end{equation}
where $g^{AB}$ is the inverse of the 2-metric normal to the 2-spheres of symmetry, i.e. of $-\dd t^2 + (\dd r + N^r(t,r)\dd t)^2$~. One then finds
\begin{equation}
    \kappa_H = \frac{1}{4M_{H}}\left[1 + (\d{M}_H - 2M^{\p}_H) \right]~, \label{eq:sg-pg}
\end{equation}
and so, eqn. \eqref{eq:metric-horizon-1} becomes
\begin{equation}
    \dd s^2|_{H} = \frac{16M_H \d{M}_H}{(1-2M^{\p}_H)^2}\kappa_H~\dd t^2 + r_H\dd \Omega^2~.
\end{equation}
A useful fact about $\kappa_H$ is that 
\begin{equation}
    \mL_{n_{-}}\theta_{+}|_{H} = -\sqrt{2}\kappa_{H}~,
\end{equation}
which can be verified by a straightforward calculation. Since $\mathcal{L}_{n_{-}}\theta_{+} < 0 $ implies that a future trapping horizon is outer in Hayward's sense \cite{hayward-bh-dynamics}, we learn that $\kappa_H > 0$ if and only if the future trapping horizon is outer. In particular, the part of the horizon we have called outer, namely $H_O$, is also outer in Hayward's sense and has positive surface gravity. 

By way of illustration, let us apply the formulae developed above to the framework for quantum-corrected dust collapse described in the preceding section. First of all, regardless of the explicit form of the mass function $M$, we know that it is bounded above (cf.~\eqref{eq:mass-bound}). Since $r_H = 2M_H$, and in particular, $M_{H_O} = M_{\text{ADM}}$, we obtain a lower bound on the mass of the black hole:
\begin{equation}
    M_{\text{ADM}} \geq \frac{\sqrt{\mu}}{2}~. \label{eq:m-adm-bound}
\end{equation}
Thus, we see that a black hole forms only if the ADM mass of the collapsing body is greater than the Planck mass ($\sim \sqrt{\hbar}\sim \sqrt{\mu}$). 

For the quantum-corrected Oppenheimer-Snyder model, we can calculate the surface gravity \eqref{eq:sg-pg} explicitly. For the outer horizon, $\kappa_{H_O} =1/4M_{\text{ADM}}$, regardless of the model of collapse. Thus, it is only the surface gravity at the inner horizon $H_I$ which will be different for different models. For the quantum-corrected Oppenheimer-Snyder model, by setting $M_{i}(t,r) = M_{H_I} =r_{H_I}/2$ (\eqref{eq:mass-inner-os}), we get 
\begin{equation}
    M_{H_I} = \frac{1}{2}\sqrt{\mu + \frac{9}{4}(t-t_0)^2}~,
\end{equation}
from which we obtain
\begin{equation}
    \kappa_{H_I} = \frac{1}{2\sqrt{\mu+ \frac{9}{4}(t-t_0)^2}}\left[1 + \frac{9}{4}\frac{(t-t_0)}{\sqrt{\mu+ \frac{9}{4}(t-t_0)^2}} \right]~.
\end{equation}
One particular approximation to this expression will be useful later. If $(t-t_0)\gg \mu$, which will be true soon after the bounce (since $\mu\ll 1$), we may turn off $\mu$ and write
\begin{equation}
    \kappa_{H_I} \approx \frac{5}{6(t-t_0)}~. \label{eq:sg-os-after-bounce}
\end{equation}
Applying the same approximation to $r_{H_I}=2M_{H_I}$, we can also write
\begin{equation}
    r_{H_I} = \frac{3}{2}(t-t_0)~. 
\end{equation}
One final aspect of the Oppenheimer-Snyder model that is worth bearing in mind is the following. The expression for $M_{H_I}$ implies that
\begin{align}
    \frac{4\d M_{H_I}}{(1-M^\p_{H_I})^2}[1-&(\d M_{H_I} -2M^\p_{H_I})] = 5\d M_{H_I}\\
    &= \frac{45}{8}\frac{t-t_0}{\sqrt{\mu + \frac{9}{4}(t-t_0)^2}}~.
\end{align}
It follows that during the collapse (rebound) phase, the metric on the horizon \eqref{eq:metric-horizon-1} has negative (positive) signature, confirming our expectation that the inner horizon should be timelike before the bounce and spacelike after the bounce.

\subsection{Minimal assumptions on quantum-corrected collapse}
It is important to emphasize that the particular framework of effective LQG and the specific example of the Oppenheimer-Snyder model were presented to illustrate concretely how the picture of quantum-corrected collapse depicted in Fig.~\ref{fig:non-sing-collapse-1} can be mathematically realized. Specific features of the idealized quantum-corrected collapse that have been presented above, such as the explicit form of the mass function and surface gravity, the speed with which the boundary of the collapsing matter moves, and so on, may turn out to be completely different if obtained from a fully consistent and satisfactory theory of quantum gravity. For the purposes of this paper, we require only the following minimal assumptions on quantum-corrected collapse in spherical symmetry, gleaned from our foregoing study of the framework of effective LQG.
\begin{enumerate}
    \item Collapse and expansion occur in a single asymptotic region, as in Fig.~\ref{fig:non-sing-collapse-1}. This means that the spacetime can be covered by generalized PG coordinates using the metric \eqref{eq:pg-metric}. 
    \item The conserved mass of the collapsing matter should be much larger than the Planck mass, $M_{\text{ADM}} \gg \sqrt{\hbar}$, consistent with \eqref{eq:m-adm-bound}. This in turn implies that the trapped region in Fig.~\ref{fig:non-sing-collapse-1} is macroscopic, i.e., $r_{H_O}\gg \sqrt{\hbar}$. 
\end{enumerate}
The justification for the first assumption is its attestation in a growing number of quantum-corrected models of gravitational collapse \cite{Saini:2014qpa, Greenwood:2008ht, Wang:2009ay, Torres:2014pea, Bambi:2013gva, Bambi:2016uda, Liu:2014kra, Husain:2021ojz, Husain2024, Wilson-Ewing:2016yan, Cipriani:2024nhx, Bojowald:2024ium, Ashtekar:2011ni, Haggard:2014rza, Han:2023wxg, Giesel:2023hys, Giesel:2023tsj, Alonso-Bardaji:2025hda, Callan:1992rs}. The second assumption is necessary because such models are to be understood in the framework of an effective theory which incorporates some, but by no means all, crucial aspects of quantum gravity. As such, there is no guarantee that these models give a fully reliable picture in the small and highly quantum region where the collapsing matter bounces \cite{Hossenfelder:2009xq}. Therefore, any analyses using such models should be performed only in regions far away from the unknown fully quantum region. This can be done only if the size of the trapping region is much larger than the quantum region where the bounce occurs.

This concludes our review of the tools required to study effective dynamical black holes. Next we review a framework in which spontaneous emission by such black holes can be conveniently described.

\section{Quantum theory of a relativistic particle}\label{sec:qm-rel-particle}
The framework to study the spontaneous emission of radiation by a black hole is that of quantum field theory on curved spacetime. To be concrete, let us take a massive Klein-Gordon scalar field on a spacetime. It satisfies the Klein-Gordon equation
\begin{equation}
    \Box \phi - \frac{m^2}{\hbar^2}\phi = 0~, \label{eq:kg}
\end{equation}
where the d'Alembertian $\Box$ is evaluated using the background metric. While a typical understanding of Klein-Gordon theory involves the formalism of quantum field theory, there is another formalism, called the proper-time formalism, which is able to describe the physics of the Klein-Gordon equation in terms of the quantum mechanics of a relativistic particle. While originally envisaged in the context of quantum electrodynamics \cite{Feynman:1950ir, Fock:1937dy, Nambu:1950rs, schwinger-1951, Stueckelberg:1941rg}, the formalism can be extended to study the quantum mechanics of a relativistic particle on a curved spacetime, as done, for instance, by Hartle and Hawking \cite{Hartle:1976tp}. To begin with, the particle will trace out a curve in spacetime which can be specified by writing down the coordinates $x^a(\la)$ of the particle as a function of a parameter $\la$ along the curve. Classically, the motion of the particle from a parameter value, say $\la =0$, to a parameter value, say $\la = \la_0$, can be described by extremizing its action functional, which is
\begin{equation}
    I[x(\la)] = \frac{1}{2}\int_{0}^{\la_0} g_{ab}\frac{\dd x^a}{\dd\la}\frac{\dd x^b}{\dd\la}\dd\la~. \label{eq:particle-action}
\end{equation}
Extremizing this action shows that $x^a(\la)$ traces out a geodesic in spacetime and $\la$ is an affine parameter along the geodesic. Thus, for a massive particle, for instance, we may take $\la$ to be proportional to the proper time of the particle. At the quantum level, one would like to associate a wave function with the particle and write down a Schrodinger equation which encodes the physics inherent in the Klein-Gordon equation. In accordance with the fact that the trajectory of the particle is described by specifying four coordinates $x^a(\la)$ as functions of a fifth parameter $\la$, we will have a wave function $\psi(x,\la)$ which depends on the given five variables. For the Schrodinger equation, one takes
\begin{equation}
    \frac{i}{\hbar}\frac{\partial \psi}{\partial\la} = -\Box\psi~. \label{eq:schr}
\end{equation}
To see that this is equivalent to the Klein-Gordon equation \eqref{eq:kg}, consider a solution $\phi$ to the Klein-Gordon equation. Then one can obtain a solution of the Schrodinger equation \eqref{eq:schr} simply by writing $\psi(x,\la) = e^{im^2\la/\hbar}\phi(x)$. Conversely, from a solution $\psi$ to the Schrodinger equation, one can obtain a solution to the Klein-Gordon equation by Fourier-transforming, i.e.
\begin{equation}
    \phi(x) = \int_{-\infty}^{\infty}\dd\la~e^{-im^2\la/\hbar}\psi(x,\la)~. \label{eq:psi-to-phi}
\end{equation}

The advantage of the Schrodinger-equation-in-parameter-time formulation is that one can now use path-integral methods to the study the quantum mechanics of a relativistic particle. To this end, we can first write down the amplitude of a particle to propagate from a point $x'$ at $\la = 0$ to a point $x$ at $\la=\la_0$, namely
\begin{align}
    F(\la_0,x,x') &= \int \mathcal{D}x(\la) \exp\left(\frac{i}{\hbar}I[x(\la)] \right)\nonumber \\
    &= \int \mathcal{D}x(\la) \exp\left(\frac{i}{2\hbar} \int_{0}^{\la_0}g_{ab}\frac{\dd x^a}{\dd \la}\frac{\dd x^b}{\dd \la}\right)~. \label{eq:prop-param}
\end{align}
Imposing the boundary condition $F(0,x,x') = \frac{1}{\sqrt{-g}}\delta^{(4)}(x-x')$, one can show \cite{Hartle:1976tp} that this parameter-dependent propagator satisfies its own Schrodinger equation,
\begin{equation}
    \frac{i}{\hbar}\frac{\partial F}{\partial \la_0} = -\frac{1}{2}\left(\Box - \frac{R}{3}\right) F~, \label{eq:schro-prop}
\end{equation}
where $R$ is the Ricci scalar on the background spacetime. Thus, the wave function $\psi$ at $(x,\la_0)$, that is to say the amplitude of arrival at the spacetime point $x$ in parameter time $\la$, can be written using superposition as
\begin{equation}
    \psi(x,\la_0) = \int \dd^4 x' ~F(\la_0,x,x')\psi(x',0)~. \label{eq:prop-to-wf}
\end{equation}
Of course, the parameter $\la$ is unobservable, and the real quantity of interest is the amplitude $K(x,x')$ to propagate from $x'$ to $x$ in \textit{any} parameter time. This amplitude can be obtained from a weighted average of $F(\la_0,x,x^\p)$ over $\la_0$ \cite{Hartle:1976tp}:
\begin{equation}
    K(x,x') = i\int_{0}^{\infty}\dd\la_0~e^{-im^2\la_0/\hbar} F(\la_0, x,x')~. \label{eq:prop-param-to-prop-kg}
\end{equation}
In words, this states that the amplitude to propagate from $x'$ to $x$ is given by the sum over all paths from $x'$ to $x$ in parameter time $\la_0$ followed by a sum over all parameter times. The sum runs over only positive $\la_0$ because $F(\la_0,x,x') = 0$ for $\la_0<0$ by construction. By virtue of \eqref{eq:schro-prop}, $K(x,x')$ satisfies
\begin{equation}
    \left(\Box - \frac{m^2}{\hbar^2}\right)K(x,x')_ = -\frac{\delta^{(4)}(x-x')}{\sqrt{-g}}~, \label{eq:prop}
\end{equation}
which shows that $\braket{x|x'}$ is indeed the propagator for Klein-Gordon theory. 

In summary, the quantum mechanics of a relativistic particle is enshrined in the three basic equations \eqref{eq:prop-param}, \eqref{eq:schro-prop} and \eqref{eq:prop-param-to-prop-kg}, and these equations can be studied in lieu of the quantum field theory of a Klein-Gordon scalar field. As we shall see, combined with some semiclassical methods, this alternative viewpoint enables one to study spontaneous emission of radiation by dynamical gravitational collapse, such as is described by the metric \eqref{eq:pg-metric}. 

\section{Spontaneous emission as tunneling}\label{sec:rad-tunneling-methods}

Particle creation by a black hole can be understood from the perspective of pair creation via quantum mechanical tunneling \cite{Vanzo:2011wq, padmanabhan-99, Parikh:1999mf, BROUT1991209, stephens-1989, Damour:1976jd, Wondrak:2023zdi, Boasso:2024ryt}. Under normal circumstances, the vacuum of a free quantum field is populated by particle-antiparticle pairs popping into existence and annihilating as soon as they are formed -- the phenomenon of virtual pair-creation. Thus, on average, an observer will detect no particles. However, in the presence of a strong potentials (e.g. electromagnetic fields, gravitational fields, etc.), some of the virtual particle-antiparticle pairs might get separated from each other, and hence can eventually be registered by a particle detector -- the phenomenon of real pair creation. The fact that this process cannot occur classically can be thought of in terms of there being a ``barrier'' which prevents particle-antiparticle pairs from being separated; quantum mechanically, however, there can be tunneling through a barrier, and thus the possibility suggests itself of understanding spontaneous emission by black holes as a tunneling phenomenon.  

So long as one is not interested in the precise proportionality factors that go into determining the tunneling probability, quantum mechanical tunneling can be studied in the semiclassical approximation. This is the regime in which the Planck constant is small but not small enough to wash away all effects of quantum theory. To make these ideas precise, consider the propagator $K(x,x')$. An approximate form of $K(x,x')$, valid to first order in $\hbar$, is given by \cite{Morette:1951zz}
\begin{equation}
    K(x,x') = N\exp \left(\frac{i}{\hbar}I[x,x'] \right)~, \label{eq:prop-param-approx}
\end{equation}
where $N$ is a normalization factor which depends on the van-Vleck-Morette determinant \cite{Morette:1951zz} and the notation $I(x,x')$ indicates that the action is to be evaluated along specific path(s) from $x'$ to $x$, namely those that satisfy the classical equations of motion. A way to state the classical equations of motion, and one we shall find the most convenient, is the Hamilton-Jacobi equation for the action. This can be verified by substituting the above equation into the propagator equation \eqref{eq:prop}; we get
\begin{equation}
    \partial_aI\partial^aI + \frac{m^2}{\hbar^2} = 0~. \label{eq:hamilton-jacobi}
\end{equation}
If we keep in mind the fact that the Lagrangian $\frac{1}{2}g_{ab}\frac{\dd x^a}{\dd\la}\frac{\dd x^b}{\dd\la}$ for a massive relativistic particle of mass $m$ is equal to $-m^2/2$, then we realize that $\partial I/\partial\la = -m^2/2$ and thus the preceding equation is nothing but the Hamilton-Jacobi equation for $I$. In other words, in \eqref{eq:prop-param-approx}, the action is to be evaluated along paths that satisfy the Hamilton-Jacobi equation. The equation \eqref{eq:prop-param-approx} is the path-integral analog of the WKB approximation to the wave function in quantum mechanics \cite{Morette:1951zz, holstein-swift-1}.   

Now, it might appear that the presence of only classical paths in \eqref{eq:prop-param-approx} precludes the possibility of using this equation to study quantum mechanical tunneling. The appearance, however, is illusory. To understand this point, it is instructive to recall how tunneling through a classically forbidden region is studied in the WKB approximation to the Schrodinger equation \cite[Chapter~VII]{Landau:1991wop}. Strictly speaking, the WKB approximation breaks down near the classical turning points on the particle's trajectory. To deal with this problem, one allows the particle's path to be complex and passes around the turning points via a contour in the complex plane. This yields an approximate form of the wave function in the classically forbidden region, which is then matched with the WKB form of the wave function in the classically allowed region(s) via certain connection formulae (see Ref.~\cite[Chapter~VII]{Landau:1991wop} for details). The essential point is that the analytic continuation around a classical turning point imparts an imaginary exponential contribution to the wavefunction; in this way, the amplitude-squared of the wave function gives a finite probability of transmission through the classically forbidden region. (For an application of this method to spontaneous emission by a Schwarzschild black hole, see \cite{Damour:1976jd, padmanabhan-99}.)

Essentially similar ideas as above can be applied to the approximate path-integral propagator \eqref{eq:prop-param-approx} in a much simpler manner \cite{holstein-swift-2}. One again allows the particle to traverse around a classical turning point in the complex plane. The action is then evaluated along this modified path and thus incurs an imaginary part. Up to a normalization factor, the evaluation of which would necessitate going beyond the semiclassical approximation, the probability $\sigma$ of the particle's following the modified path can be directly obtained from \eqref{eq:prop-param-approx} by writing
\begin{equation}
    \sigma = |K(x,x')|^2 \sim \exp\left(-\frac{2}{\hbar}\text{Im}[I(x,x')]\right)~. \label{eq:sc-prob}
\end{equation}
This approach is much simpler than the one using the WKB approximation to the Schrodinger equation, since the need to perform rather intricate derivations of the precise connection formulae in the latter approach is now obviated. The only nontrivial part now is the determination of the complex path followed by the particle in the classically forbidden region. While this is a nontrivial task and carrying it out differs from case to case\footnote{See, for instance, the difficulties involved in the two examples studied in Ref.~\cite{holstein-swift-2}.}, in the context of spontaneous emission by black holes, there is an elegant method which allows one to obtain the required imaginary contribution to the action without considerable effort. This method is based on the Hamilton-Jacobi equation and an energy-conservation argument suggested by the heuristic picture of spontaneous emission as pair creation by strong fields that we outlined at the start of this section. We shall now describe the application of this method to non-singular gravitational collapse (Fig.~\ref{fig:non-sing-collapse-1}). 

%since we are in the semiclassical regime, where such quantum effects as tunneling are present, there will be paths corresponding to such phenomena. If one can identify these, one can then use \eqref{eq:prop-param-to-prop-kg} to find the probability amplitude to tunnel along these paths. In the presence of a black hole, some of these tunneling paths will correspond to the separation of particle-antiparticle pairs by the black hole, and so evaluating the action along such paths will inform us about spontaneous emission. This is the basic idea that will now be spelled out in detail. 

To start with, let us take a closer look at the Hamilton-Jacobi equation \eqref{eq:hamilton-jacobi} holds. For convenience, we will consider massless particles ($m=0$). This simplification is justified by the fact that it is sufficient to obtain the correct probability of emission in all conventionally studied examples, such as static black holes, rotating black holes, and late-time classical gravitational collapse \cite{Vanzo:2011wq}. It is also sufficient to focus attention on radial paths (i.e., zero angular momentum). With these assumptions, the Hamilton-Jacobi equation \eqref{eq:hamilton-jacobi} for the metric \eqref{eq:pg-metric} reads
\begin{equation}
    [1-(N^r)^2](\partial_rI)^2 + (\pd_t I)^2 + 2N^r\pd_r I\pd_t I = 0~, 
\end{equation}
and since classical radial paths are radial null geodesics, we also have
\begin{equation}
    -[1-(N^r)^2]\left(\frac{\dd t}{\dd\la}\right)^2 + \left(\frac{\dd r}{\dd\la}\right)^2 + 2N^r\frac{\dd t}{\dd \la}\frac{\dd r}{\dd\la} = 0~.
\end{equation}
These equations are equivalent to
\begin{align}
    \frac{\dd r}{\dd t} &= \pm (1\mp N^r) = \theta_{\pm}~, \label{eq:null-speed}\\
     p_r &= \mp\frac{p_t}{1\mp N^r} = -\frac{p_t}{\theta_{\pm}}~ \label{eq:null-mom}~,
\end{align}
where we have used \eqref{eq:theta-pm} and also replaced $\pd_a I$ with the momentum $p_a$ conjugate to $x^a$, i.e., 
\begin{equation}
    p_a \equiv \frac{\pd L}{\pd(\dd x^a/\dd\la)} = g_{ab}\frac{\dd x^b}{\dd\la} = \partial_aI~,
\end{equation}
where the last equality follows from the Hamilton-Jacobi equation and the equation for null geodesics.

The equation \eqref{eq:null-speed} gives the speed along ingoing ($\theta_{-}$) and outgoing ($\theta_{+}$) radial null geodesics, while \eqref{eq:null-mom} gives the corresponding radial momentum. The latter can be written in a more illuminating manner in terms of a coordinate-invariant energy $\om(t,r)$ associated with a trajectory. In a general spacetime, there is no notion of the energy of a particle. However, in spherical symmetry, the Kodama vector field, introduced in Sec.~\ref{sec:bh-dynamics}, gives a preferred flow of time which reduces to that of static observers (i.e. Killing time) in the static limit. This vector field can be used to define the invariant energy $\om(t,r)$ \cite{Vanzo:2011wq, Hayward:2008jq, DiCriscienzo:2009kun},

\begin{equation}
    \om \equiv -p_aK^a = -p_t~. 
\end{equation}
Asymptotically, $K^a$ becomes a unit timelike vector field, hence this definition corresponds to the usual definition of particle energy at asymptotic infinity. In the rest of the spacetime, one can give the following physical meaning to $\om$ as defined above. Consider a static observer with proper time $\tau$ outside the trapping horizon. The 4-velocity $u^a = (\dd t/\dd\tau, 0,0,0)$ of such an observer can be written down as
\begin{equation}
    u^a \equiv \frac{\dd x^a}{\dd \tau} = \frac{\dd x^a}{\dd t}\frac{\dd t}{\dd\tau} = -\frac{K^a}{\theta_{+}\theta_{-}}~.
\end{equation}
Thus, the energy of a particle with momentum $p^a$ as measured by this observer must be
\begin{equation}
    \om_{\text{st-obs}} = -u^ap_a = -\frac{\om}{\theta_{+}\theta_{-}}~.
\end{equation}
As we saw in Sec.~\ref{sec:bh-dynamics}, outside the horizon, $\theta_{+} > 0$ and $\theta_{-} < 0$. Thus, $\om$ must have the same sign as $\om_{\text{st-obs}}$ outside the trapping horizon. Therefore, we may simply think of $\om$ as the particle energy, keeping in mind that it is actually proportional to the energy measured by a static observer; we will call $\om$ the Kodama energy. With this understood, we now write the radial momentum \eqref{eq:null-mom} along a particle trajectory as
\begin{equation}
    p_r = \frac{\om}{\theta_{\pm}}~. 
\end{equation}
To summarize, the trajectory of a particle is characterized by specifying its radial speed \eqref{eq:null-speed}, Kodama energy $\om$ and radial momentum $p_r$, the latter two being related via the preceding equation. 

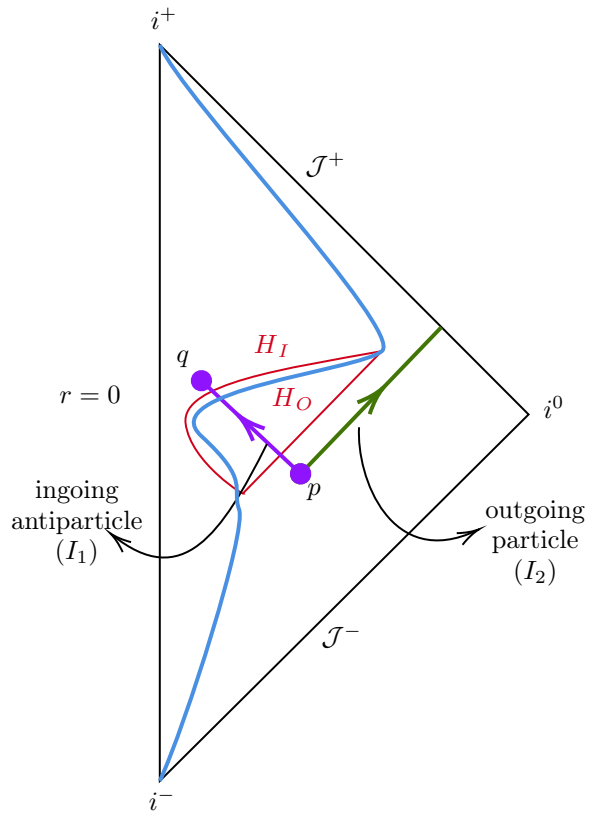
\begin{figure}[ht]
   
   \tikzset{every picture/.style={line width=0.75pt}} %set default line width to 0.75pt        

\begin{tikzpicture}[x=0.75pt,y=0.75pt,yscale=-1,xscale=1]
%uncomment if require: \path (0,573); %set diagram left start at 0, and has height of 573

%Shape: Triangle [id:dp4190537488403724] 
\draw   (565,308.04) -- (379,493) -- (379,121.5) -- cycle ;
%Straight Lines [id:da7879793296761517] 
\draw [color={rgb, 255:red, 208; green, 2; blue, 27 }  ,draw opacity=1 ]   (491,276.07) -- (421.05,347.76) ;
%Curve Lines [id:da9396701303205112] 
\draw [color={rgb, 255:red, 208; green, 2; blue, 27 }  ,draw opacity=1 ]   (421.05,347.76) .. controls (423.47,350.28) and (394.43,332.65) .. (392.01,312.5) .. controls (389.59,292.35) and (442.83,284.8) .. (491,276.07) ;

%Straight Lines [id:da8900187555256068] 
\draw [color={rgb, 255:red, 65; green, 117; blue, 5 }  ,draw opacity=1 ][line width=1.5]    (521,264) -- (450,338) ;
\draw [shift={(450,338)}, rotate = 133.81] [color={rgb, 255:red, 65; green, 117; blue, 5 }  ,draw opacity=1 ][fill={rgb, 255:red, 65; green, 117; blue, 5 }  ,fill opacity=1 ][line width=1.5]      (0, 0) circle [x radius= 4.36, y radius= 4.36]   ;
\draw [shift={(491.59,294.65)}, rotate = 133.81] [color={rgb, 255:red, 65; green, 117; blue, 5 }  ,draw opacity=1 ][line width=1.5]    (14.21,-4.28) .. controls (9.04,-1.82) and (4.3,-0.39) .. (0,0) .. controls (4.3,0.39) and (9.04,1.82) .. (14.21,4.28)   ;
%Straight Lines [id:da751459919037825] 
\draw [color={rgb, 255:red, 144; green, 19; blue, 254 }  ,draw opacity=1 ][line width=1.5]    (400,291) -- (450,338) ;
\draw [shift={(450,338)}, rotate = 43.23] [color={rgb, 255:red, 144; green, 19; blue, 254 }  ,draw opacity=1 ][fill={rgb, 255:red, 144; green, 19; blue, 254 }  ,fill opacity=1 ][line width=1.5]      (0, 0) circle [x radius= 4.36, y radius= 4.36]   ;
\draw [shift={(418.59,308.47)}, rotate = 43.23] [color={rgb, 255:red, 144; green, 19; blue, 254 }  ,draw opacity=1 ][line width=1.5]    (14.21,-4.28) .. controls (9.04,-1.82) and (4.3,-0.39) .. (0,0) .. controls (4.3,0.39) and (9.04,1.82) .. (14.21,4.28)   ;
\draw [shift={(400,291)}, rotate = 43.23] [color={rgb, 255:red, 144; green, 19; blue, 254 }  ,draw opacity=1 ][fill={rgb, 255:red, 144; green, 19; blue, 254 }  ,fill opacity=1 ][line width=1.5]      (0, 0) circle [x radius= 4.36, y radius= 4.36]   ;
%Curve Lines [id:da6589031400654984] 
\draw    (479.5,314.5) .. controls (478.02,339.75) and (495.64,387.04) .. (538.69,366.64) ;
\draw [shift={(540,366)}, rotate = 153.43] [color={rgb, 255:red, 0; green, 0; blue, 0 }  ][line width=0.75]    (10.93,-3.29) .. controls (6.95,-1.4) and (3.31,-0.3) .. (0,0) .. controls (3.31,0.3) and (6.95,1.4) .. (10.93,3.29)   ;
%Curve Lines [id:da9551060201174512] 
\draw    (433,323) .. controls (410.23,368.54) and (390.4,397.42) .. (355.08,368.88) ;
\draw [shift={(354,368)}, rotate = 39.81] [color={rgb, 255:red, 0; green, 0; blue, 0 }  ][line width=0.75]    (10.93,-3.29) .. controls (6.95,-1.4) and (3.31,-0.3) .. (0,0) .. controls (3.31,0.3) and (6.95,1.4) .. (10.93,3.29)   ;
%Curve Lines [id:da047880391489928176] 
\draw [color={rgb, 255:red, 74; green, 144; blue, 226 }  ,draw opacity=1 ][line width=1.5]    (379,121.5) .. controls (389.71,147.85) and (502.81,266.15) .. (491,276.07) .. controls (479.19,286) and (375,296) .. (400,319) .. controls (425,342) and (415,345) .. (419,356) .. controls (423,367) and (392,465) .. (379,493) ;

% Text Node
\draw (372,495.4) node [anchor=north west][inner sep=0.75pt]    {$i^{-}$};
% Text Node
\draw (373,101.4) node [anchor=north west][inner sep=0.75pt]    {$i^{+}$};
% Text Node
\draw (571,297.4) node [anchor=north west][inner sep=0.75pt]    {$i^{0}$};
% Text Node
\draw (459,412.4) node [anchor=north west][inner sep=0.75pt]    {$\mathcal{J}^{-}$};
% Text Node
\draw (451,174.4) node [anchor=north west][inner sep=0.75pt]    {$\mathcal{J}^{+}$};
% Text Node
\draw (327,292.4) node [anchor=north west][inner sep=0.75pt]    {$r=0$};
% Text Node
\draw (452,341.4) node [anchor=north west][inner sep=0.75pt]    {$p$};
% Text Node
\draw (539,350) node [anchor=north west][inner sep=0.75pt]   [align=left] {\begin{minipage}[lt]{41.86pt}\setlength\topsep{0pt}
\begin{center}
outgoing\\particle\\($I_2$)
\end{center}

\end{minipage}};
% Text Node
\draw (301,340) node [anchor=north west][inner sep=0.75pt]   [align=left] {\begin{minipage}[lt]{52.05pt}\setlength\topsep{0pt}
\begin{center}
ingoing \\antiparticle\\($I_1$)
\end{center}

\end{minipage}};
% Text Node
\draw (434,293.4) node [anchor=north west][inner sep=0.75pt]    {$\textcolor[rgb]{0.82,0.01,0.11}{H}\textcolor[rgb]{0.82,0.01,0.11}{_{O}}$};
% Text Node
\draw (425,266.4) node [anchor=north west][inner sep=0.75pt]    {$\textcolor[rgb]{0.82,0.01,0.11}{H}\textcolor[rgb]{0.82,0.01,0.11}{_{I}}$};
% Text Node
\draw (386,274.4) node [anchor=north west][inner sep=0.75pt]    {$q$};

\end{tikzpicture}

    \caption{A typical particle creation event. A particle-antiparticle pair is separated by the background gravitational field near the horizon at $r=p$. The particle travels outwards towards future null infinity. The antiparticle follows an ingoing path, eventually annihilating a shell of matter either before reaching the inner horizon ($q=r_{H_I}+\epsilon$) or at the inner horizon ($q = r_{H_I} -\epsilon$).}
    \label{fig:pair-creation}
\end{figure}

We now have all the ingredients in place to find the probability of pair creation by the kind of black hole depicted in Fig.~\ref{fig:non-sing-collapse-1}. Consider a pair creation event at $p = r_{H_O}+\epsilon$ for some small $\epsilon > 0$ (Fig.~\ref{fig:pair-creation}). We envisage the gravitational field in the region to be strong enough to separate the created particle and antiparticle such that there is an eventual flux of outgoing particles at future null infinity. Since particles/antiparticles must follow null geodesics in the semiclassical approximation, the particle must follow an outgoing null geodesic, whereas the antiparticle must follow an ingoing geodesic. Therefore,
\begin{align}
    \frac{\dd r}{\dd t} &= \theta_{+} \quad \text{(particle)},\\
    \frac{\dd r}{\dd t} &= \theta_{-} \quad \text{(antiparticle)}.
\end{align}
This fixes the the radial speed along particle and antiparticle paths. Next we fix the radial momentum along each. Suppose that the particle has energy $\om$; hence the antiparticle will have energy $-\om$. Since an observer intersecting the path of the outgoing particle must observe a positive flux of energy and momentum, we must have that $\om > 0$ and hence ($\theta_{+}>0$ in the non-trapped region)
\begin{equation}
    p_r = \frac{\om}{\theta_{+}} \quad \text{(particle)}
\end{equation}
along the outgoing particle's trajectory. By energy and momentum conservation, the momentum along the antiparticle's trajectory must therefore be
\begin{equation}
    p_r = -\frac{\om}{\theta_{+}} \quad \text{(antiparticle)}
\end{equation}

%Now, for particles, $\om > 0$ (asymptotically), whereas for antiparticles $\om < 0$ (asymptotically). Thus, for instance, along an outgoing path, although $\dd r/\dd t = \theta_{\pm}$ depending on whether the path corresponds to a particle or an antiparticle, $p_r = \om/\theta_{+}$ gives the correct sign for the radial momentum in either case. This fact will be crucial in what follows. 

%All ingredients are now in place to find the amplitude for the separation of particle-antiparticle pairs by the black hole via quantum mechanical tunneling. Fig.~\ref{fig:pair-creation} shows a typical pair creation event. A particle-antiparticle pair is created in the vicinity of the future trapping horizon at $r=p$ and then separated by the background gravitational field. The particle travels outwards towards future null infinity, since $\dd r/\dd t = \theta_{+} > 0$ along an outgoing trajectory in the non-trapped region. The antiparticle can be imagined to start from future null infinity. It then travels backward in time along an ingoing path until it reaches $r = 0$, whence it reflects and follows an outgoing trajectory. Since $\dd r/\dd t = \theta_{-} < 0$ and $\dd t < 0$ along an outgoing antiparticle path, the antiparticle travels right across the trapped region via a null path in the direction of increasing $r$ until it reaches the pair creation event at $r = p$. 

What should be the eventual fate of the particle and the antiparticle? The particle reaches future null infinity. The ingoing antiparticle, on the other hand, must be absorbed by the the collapsing matter at, say, $q$. Physically this can be understood as follows. Matter undergoing gravitational collapse in spherical symmetry can be thought of as a collection of infinitely many spherical shells, each labeled by a parameter $s$ and falling under the weight of all the shells inside it (see Ref.~\cite{Bondi:1947fta} for the classical and Refs.~\cite{Giesel:2023hys, Fazzini:2023ova} for the effective case). That is, the equations of motion for the shells decouple, each shell's dynamics being given by its radius $r(t,s)$ as a function of time. In a simplified model (see, e.g., Ref.~\cite{ford-parker-78}), we may assign an invariant proper mass $\mu$ to each shell such that the total mass of all shells in an interval $[0,s]$, where $0$ labels the inner most shell, is $M(s) = \mu\int_0^{s} ~\dd s$\footnote{Here $M$ is the same mass parameter that occurs in the metric \eqref{eq:pg-metric}. In the $(t,s)$ coordinates, this parameter is independent of $t$ \cite{Bondi:1947fta}. This assumption and with it the whole picture of independent shells falling under their gravity breaks down if the falling shells cross, which can happen for quite generic initial data \cite{Fazzini:2023ova}. We will speculate in the concluding section on how the physical picture we have outlined changes if shell crossings do occur.}. Now, in a more realistic model of pair-creation than we have described above, the created particle/antiparticle will have a proper mass, say $m$. Since in this case,
\begin{equation}
    g_{ab}\frac{\dd x^a}{\dd\la}\frac{\dd x^b}{\dd\la} = -m^2~,
\end{equation}
one must have $\dd \tau = \pm m\dd\la$, where $\tau$ is the proper time along the particle/antiparticle trajectory, and the plus (minus) sign refers to a particle (anti-particle). Following Stueckelberg \cite{Stueckelberg:1941rg} and Feynman \cite{feynman-49-positrons}, this invites us to think of the ingoing antiparticle as a particle moving backward in time, which particle can be thought of as having come from the collapsing matter. In other words, we can think of an ingoing anti-particle of proper mass $m$ (but having negative energy $-\om$ according to a static observer outside the black hole) being absorbed by the collapsing matter at $q$ as the collapsing matter having shed off spherical shells of total mass $m$ in the past at $p$ in the form of an outgoing particle of proper mass $m$ and energy $\om$. In the process, the black hole proper mass $M(s_o)$, where $s_o$ labels the shell at the boundary of the collapsing matter, reduces to $M(s_o)-m$. 

Now, the absorption of the ingoing antiparticle must occur at the outer boundary of the collapsing matter (blue line in Figs.~\ref{fig:non-sing-collapse-1} and \ref{fig:pair-creation}). In non-singular gravitational collapse \cite{Husain:2021ojz, Husain:2022gwp, Husain2024}, this boundary either follows the trajectory of the inner horizon $H_I$ or always lies above it. Therefore, the absorption would occur either before the ingoing antiparticle reaches the inner horizon or precisely at the point of intersection of the two. We accommodate the former (latter) possibility by setting the radial coordinate $q$ at the absorption event to $r_{H_I} + \epsilon$ ($r_{H_I}-\epsilon$); $\epsilon > 0$ will of course be set to zero at the end of our calculations.

Having identified the particle-antiparticle paths and the parameters characterizing them, we can now evaluate the particle action along them and subsequently use \eqref{eq:sc-prob} to find the probability of the pair-creation process described above. First of all, the action along any radial path can be written down as
\begin{equation}
    I = \int \partial_aI\dd x^a = \int (p_t\dd t + p_r \dd r) = \int (-\ep \om \dd t  + p_r\dd r)~, \label{eq:action-expanded}
\end{equation}
where $\ep = \pm$ according to whether the path over which the integral is evaluated corresponds to a particle or an antiparticle, respectively. Secondly, we can divide the integral into two parts corresponding to the two parts into which the entire trajectory of the particle-antiparticle pair is divided in Fig.~\ref{fig:pair-creation}. Thus,
\begin{equation}
    I =  \underbrace{\int_{q}^{p} (\cdots)}_{I_1}  + \underbrace{\int_{p}^{\infty}(\cdots)}_{I_2}~.
\end{equation}
Take first $I_2$. It refers to the outgoing particle, along which we have established that
\begin{equation}
    \frac{\dd r}{\dd t} = \theta_{+}~, \quad p_r = \frac{\om}{\theta_{+}}~.
\end{equation}
In the non-trapped region, these expressions are all well-defined, and substituting them into \eqref{eq:action-expanded} yields $I_2 = 0$~.

\begin{figure}[ht]
    \centering

\tikzset{every picture/.style={line width=0.5pt}} %set default line width to 0.75pt        

\begin{tikzpicture}[x=0.5pt,y=0.5pt,yscale=-1,xscale=1]
%uncomment if require: \path (0,568); %set diagram left start at 0, and has height of 568

%Straight Lines [id:da5405291925736105] 
\draw    (236,288) -- (666,287) ;
\draw [shift={(668,287)}, rotate = 179.87] [color={rgb, 255:red, 0; green, 0; blue, 0 }  ][line width=0.75]    (10.93,-3.29) .. controls (6.95,-1.4) and (3.31,-0.3) .. (0,0) .. controls (3.31,0.3) and (6.95,1.4) .. (10.93,3.29)   ;
%Straight Lines [id:da7269019546152083] 
\draw    (314,436) -- (313.01,203) ;
\draw [shift={(313,201)}, rotate = 89.76] [color={rgb, 255:red, 0; green, 0; blue, 0 }  ][line width=0.75]    (10.93,-3.29) .. controls (6.95,-1.4) and (3.31,-0.3) .. (0,0) .. controls (3.31,0.3) and (6.95,1.4) .. (10.93,3.29)   ;
%Shape: Arc [id:dp5171008948646074] 
\draw  [draw opacity=0] (458.99,281.06) .. controls (459,281.37) and (459,281.69) .. (459,282) .. controls (459,298.57) and (445.57,312) .. (429,312) .. controls (412.43,312) and (399,298.57) .. (399,282) .. controls (399,282) and (399,282) .. (399,282) -- (429,282) -- cycle ; \draw  [color={rgb, 255:red, 74; green, 144; blue, 226 }  ,draw opacity=1 ] (458.99,281.06) .. controls (459,281.37) and (459,281.69) .. (459,282) .. controls (459,298.57) and (445.57,312) .. (429,312) .. controls (412.43,312) and (399,298.57) .. (399,282) .. controls (399,282) and (399,282) .. (399,282) ;  
%Shape: Arc [id:dp16736734876405457] 
\draw  [draw opacity=0] (580,282) .. controls (580,282) and (580,282) .. (580,282) .. controls (580,298.57) and (566.57,312) .. (550,312) .. controls (533.43,312) and (520,298.57) .. (520,282) .. controls (520,282) and (520,282) .. (520,282) -- (550,282) -- cycle ; \draw  [color={rgb, 255:red, 74; green, 144; blue, 226 }  ,draw opacity=1 ] (580,282) .. controls (580,282) and (580,282) .. (580,282) .. controls (580,298.57) and (566.57,312) .. (550,312) .. controls (533.43,312) and (520,298.57) .. (520,282) .. controls (520,282) and (520,282) .. (520,282) ;  
%Straight Lines [id:da16283232303812478] 
\draw [color={rgb, 255:red, 74; green, 144; blue, 226 }  ,draw opacity=1 ]   (250,282) -- (399.01,282.94) ;
%Straight Lines [id:da6600290143403509] 
\draw [color={rgb, 255:red, 74; green, 144; blue, 226 }  ,draw opacity=1 ]   (580,282) -- (631,282.5) ;
%Straight Lines [id:da9901359659288868] 
\draw [color={rgb, 255:red, 74; green, 144; blue, 226 }  ,draw opacity=1 ]   (459,282) -- (520.01,282.94) ;
%Shape: Arc [id:dp34268627049066425] 
\draw  [draw opacity=0] (631,282.5) .. controls (631,282.7) and (631,282.9) .. (631,283.1) .. controls (631,357.6) and (545.71,418) .. (440.5,418) .. controls (335.29,418) and (250,357.6) .. (250,283.1) .. controls (250,282.9) and (250,282.7) .. (250,282.5) -- (440.5,283.1) -- cycle ; \draw  [color={rgb, 255:red, 74; green, 144; blue, 226 }  ,draw opacity=1 ] (631,282.5) .. controls (631,282.7) and (631,282.9) .. (631,283.1) .. controls (631,357.6) and (545.71,418) .. (440.5,418) .. controls (335.29,418) and (250,357.6) .. (250,283.1) .. controls (250,282.9) and (250,282.7) .. (250,282.5) ;  
\draw  [color={rgb, 255:red, 74; green, 144; blue, 226 }  ,draw opacity=1 ] (452.87,427.77) .. controls (447.8,422.62) and (442.73,419.52) .. (437.65,418.49) .. controls (442.73,417.46) and (447.8,414.37) .. (452.87,409.21) ;
\draw  [color={rgb, 255:red, 74; green, 144; blue, 226 }  ,draw opacity=1 ] (433.51,304.3) .. controls (440.14,305.69) and (446.03,305.58) .. (451.18,303.97) .. controls (446.76,307.07) and (443.07,311.67) .. (440.13,317.76) ;
\draw  [color={rgb, 255:red, 74; green, 144; blue, 226 }  ,draw opacity=1 ] (557.75,303.78) .. controls (564.97,303.44) and (570.66,301.73) .. (574.81,298.64) .. controls (572.19,303.1) and (571.11,308.94) .. (571.57,316.16) ;
%Straight Lines [id:da5438639698331423] 
\draw    (549,288) -- (558,311) ;
%Straight Lines [id:da03992445708933712] 
\draw    (430,288) -- (422,311) ;

% Text Node
\draw (646,292.4) node [anchor=north west][inner sep=0.75pt]    {$\text{Re}( r)$};
% Text Node
\draw (277,175.4) node [anchor=north west][inner sep=0.75pt]    {$\text{Im}( r)$};
% Text Node
\draw (418,265.4) node [anchor=north west][inner sep=0.75pt]    {$r_{H_{I}}$};
% Text Node
\draw (540,264.4) node [anchor=north west][inner sep=0.75pt]    {$r_{H_{O}}$};
% Text Node
\draw (589,290.4) node [anchor=north west][inner sep=0.75pt]    {$p$};
% Text Node
\draw (298,289.4) node [anchor=north west][inner sep=0.75pt]    {$0$};
% Text Node
\draw (414,288.4) node [anchor=north west][inner sep=0.75pt]    {$\epsilon $};
% Text Node
\draw (538,288.4) node [anchor=north west][inner sep=0.75pt]    {$\epsilon $};
% Text Node
\draw (521,404.4) node [anchor=north west][inner sep=0.75pt]    {$C_{3}$};
% Text Node
\draw (390,310.4) node [anchor=north west][inner sep=0.75pt]    {$C_{1}$};
% Text Node
\draw (514,310.4) node [anchor=north west][inner sep=0.75pt]    {$C_{2}$};

\end{tikzpicture}
    \caption{Choice of contour in evaluating $I_1$.}
    \label{fig:contour}
\end{figure}
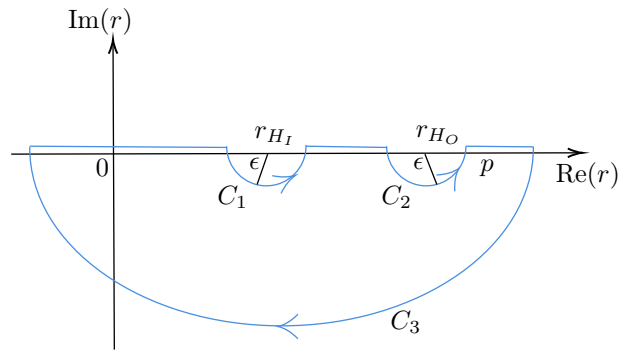

The evaluation of $I_1$ yields a more interesting result. It corresponds to the ingoing antiparticle, along which
\begin{equation}
    \frac{\dd r}{\dd t} = \theta_{-}~, \quad p_r = -\frac{\om}{\theta_{+}}~.
\end{equation}
Therefore,
\begin{equation}
    I_1 = \int_{q}^{p} \om\left(\frac{1}{\theta_{-}} - \frac{1}{\theta_{+}}\right) \dd r = 2\int_{q}^{p}\frac{\om}{\theta_{+}\theta_{-}}\dd r~, 
\end{equation}
where the last equality follows from $\theta_{\pm} = \pm(1\mp N^r)$. At the trapping horizon $H = H_{O} \cup H_{I}$, $\theta_{+} = 0$, and the antiparticle crosses it at least once. If it is absorbed by the collapsing matter before reaching the inner horizon, then it crosses the trapping horizon only once at $H_{O}$, but if it is absorbed by the collapsing matter at the location of the inner horizon, then it crosses the trapping horizon both at $H_O$ and $H_I$. In either case, since $\theta_{+} = 0$, the integrand above has pole(s) along the path of the antiparticle. These are the classical turning points that we alluded to in our discussion of semiclassical tunneling. Thus, as we anticipated then, we allow $r$ to be complex and go around the pole by deforming the contour of integration. Here one has to make a choice of whether to close the contour in the upper half complex plane or the lower half complex plane. We shall close the contour in the lower half complex plane (Fig.~\ref{fig:contour}). The rationale for this choice is the reduction of our result for the probability of spontaneous emission to the Hawking result under appropriate assumptions, as will be seen below. Hence we can write, using the residue theorem \cite{arfken-2013},
\begin{equation}
    \lim_{\ep\to 0}\left[\int_{-\infty}^{r_{H_I}-\ep}+\Delta\int_{C_1}+\underbrace{\int_{r_{H_I}+\ep}^{r_{H_O}-\ep}}_{I_1} + \int_{C_2} +\int_{C_3} \right] = 0 ~,
\end{equation}
where $\De = \{0,1\}$, depending on whether the antiparticle is absorbed at $q=r_{H_I}+\ep$ or $q=r_{H_I}-\ep$, respectively. 
The preceding equation can be rewritten more illustratively as\footnote{Strictly speaking, because we are working in spherical coordinates, the integrand is not defined for $r<0$. But the term $\int_{-\infty}^{r_{H_I}-\ep}$ necessitates that we specify the integrand on the negative real axis too. To do this, we may make any choice as long as $\om$ and $M$ are real for $r<0$. In this we are guided by the expectation that this term should not give an imaginary contribution, as is the case for a Schwarzschild black-hole, for which the integrand is $-2\om_{\infty}/(1-2M/r)$, which is real even if $r<0$. I thank Irfan Javed for drawing my attention to this subtle point.} 
\begin{align*}
    I_1 &= -\De\int_{C_1} - \int_{C_2} + \text{ real terms}\\
    &=-i\pi\sum_{\De r_{H_I}, r_{H_O}}\text{Res}\left(\frac{2\om}{\theta_{+}\theta_{-}}\right) + \text{real terms}~,
\end{align*}
the second line being an application\footnote{In conventional accounts of Hawking radiation as tunneling \cite{Vanzo:2011wq}, it is customary to use Feynman's $i\ep$ prescription to evaluate the principal value of the action along the classically forbidden trajectory. We have avoided this route in favor of the standard treatment of principal values found in mathematical physics textbooks (e.g., Ref. \cite{arfken-2013}) on account of the latter's greater clarity when the residues of the integrand cannot be found by mere inspection and instead one has to rely on a Laurent expansion.} of the calculus of residues to poles on the real line \cite{arfken-2013}. To evaluate the residues, we can expand the integrand near the horizon along the path of the ingoing particle, i.e., along $\dd r/\dd t = \theta_{-}$. \cite{Vanzo:2011wq}. Writing $\delta r = r-r_H$, we have
\begin{align}
    1-N^r = \theta_{+} &\simeq \d{\theta}_{+H}\delta t + \theta^{\p}_{+H}\delta r + O(\delta r^2)\nn \\
    &=\left[-\frac{\d{\theta}_{+}}{(1+N^r)} + \theta_{+}^{\p}\right]_{H}\delta r  + O(\delta r^2) \nn \\
    &= -\frac{1}{2}(\d{\theta}_{+} - 2\theta_{+}^{\p})_{H}\delta r  + O(\delta r^2)\nn \\
    &= \kappa_H \delta r + O(\delta r^2) \label{eq:theta-first-order}
\end{align}
Here we have used \eqref{eq:sg-pg} in the last line. Thus, to first order in $\delta r$,
\begin{equation}
    \frac{2\om}{\theta_{+}\theta_{-}} \simeq -\frac{\omega_H}{\kappa_H}\frac{1}{r-r_H}~,
\end{equation}
where $\om_H$ is the Kodama energy evaluated at the trapping horizon. Therefore,
\begin{equation}
    \text{Res}\left(\frac{2\om}{\theta_{+}\theta_{-}}\right) = -\frac{\om_H}{\kappa_H}~, \label{eq:residue}
\end{equation}
which implies that
\begin{equation}
    \text{Im}(I_1) = \De\frac{\pi\om_{H_I}}{\kappa_{H_I}} + \frac{\pi\om_{H_O}}{\kappa_{H_O}}~. 
\end{equation}
From \eqref{eq:sc-prob}, we can then conclude that the probability $\sigma$ of pair creation via the tunneling path in Fig.~\ref{fig:pair-creation} is given by
\begin{equation}
    \sigma\sim \exp\left(-\De\frac{2\pi\om_{H_I}}{\hbar\kappa_{H_I}} -\frac{2\pi\om_{H_O}}{\hbar\kappa_{H_O}} \right)~. \label{eq:prob-emission}
\end{equation}

In the classical case of a Schwarzschild spacetime, which is characterized by an eternal event horizon, one will only have the term with $H_{O}$, which will then denote the event horizon. Furthermore, in a static spherically symmetric spacetime, the Kodama vector field reduces to the timelike Killing vector field. Therefore, $\om = -p_aK^a$ is conserved and equals the energy $\om_{\infty}$ of a particle as measured by an observer at infinity (since $K^a$ becomes a unit timelike vector field asymptotically). And finally, one can read off the surface gravity from \eqref{eq:sg-pg}. The preceding formula then reduces to
\begin{equation}
    \sigma \sim \exp\left( -8\pi M_S~\om_{\infty}/\hbar \right)~, \label{eq:prob-emission-schw}
\end{equation}
where $M_S$ is the mass of the Schwarzschild black hole. This is consistent with the well-established probability of Hawking radiation in a Schwarzschild spacetime. This justifies our choice of the contour above. Had we chosen to close the contour in the upper half complex plane, we would have picked up a minus sign in applying the residue theorem, leading to a result contradicting Hawking's. For other justifications for picking the correct contour, see \cite{padmanabhan-99,Vanzo:2011wq, Parikh:1999mf}. 

For the sake of clarity, the steps leading up to \eqref{eq:prob-emission} are summarized below.
\begin{enumerate}
    \item The Klein-Gordon equation on a spacetime can be described by the path integral of a relativistic particle using the proper-time formalism \cite{Feynman:1950ir, Hartle:1976tp}.
    \item In this formalism, the propagator for a relativistic particle is given by $\braket{x|x'} = \int d\la_0 ~e^{-im^2\la_0}F(\la_0,x,x')$, where $F(\la_0,x,x')$ is the amplitude of propagation from $x'$ to $x$ in parameter time $\la_0$.
    \item In the semiclassical approximation, $K(x,x') \sim e^{iI/\hbar}$, where the action $I$ is to be evaluated along paths for which the Hamilton-Jacobi equation holds.
    \item The paths for which the Hamilton-Jacobi equation holds are null. Tunneling involves traversing classically forbidden regions using complex paths (Refs.~\cite[Chapter~VII]{Landau:1991wop} and  \cite{holstein-swift-2}). 
    \item Identify a typical tunneling path for a particle-antiparticle pair created just outside the horizon (Fig.~\ref{fig:pair-creation}). 
    \item Evaluate the particle action along the tunneling path. The choice of contour is dictated by the requirement that the tunneling probability should have the correct Schwarzschild limit. 
\end{enumerate}

Notice that in implementing step 5, we chose for the emitted particle a path which is close to the outer horizon (Fig.~\ref{fig:pair-creation}). This choice has a similar effect as the late-time approximation involved in the conventional calculation of Hawking radiation \cite{Hawking:1975vcx}. For in that calculation, the late-time approximation enables one to calculate the Bogoliubov coefficients between the late-time and early-time modes by solving the Klein-Gordon equation near the horizon. These coefficients in turn enter into the amplitude for there being particles at future null infinity at late times given that there were no particles incoming from past null infinity at early times. Somewhat similarly, the choice of a near-horizon path in the calculation above enables one to evaluate the residue \eqref{eq:residue} and hence the imaginary part of the action, which then determines the probability of spontaneous emission. It is in this sense that the result \eqref{eq:prob-emission} is the quantum-corrected analog of the classical late-time result found by Hawking.

However, as can be seen in Fig.~\ref{fig:pair-creation}, in the quantum-corrected geometry, the outgoing particle whose path snuggles the outer horizon does not reach future null infinity at late times. There might well be particles emitted by the dynamical geometry that reach future null infinity after $u=u_0$, where $u_0$ is the retarded time at which the outgoing particle in Fig.~\ref{fig:pair-creation} reaches future infinity. Now the spectra of these particles might have other corrections due to the highly nontrivial spacetime geometry to the future of $u_0$. But studying these corrections would require a knowledge of the precise details of the geometry in regions where the effective description of quantum-corrected collapse that we have employed may not be reliable (e.g., the bounce region). Therefore, our calculation sheds light only on any potential corrections to the spectrum of emitted particles which reach future infinity near $u_0$. These paths are close to the outer horizon, and as such enable us to abstract away the effect of the classical late-time approximation which is responsible for producing a thermal spectrum in the conventional case.

Before closing this section, a historical note is in order. The Hamilton-Jacobi method, encapsulated in \eqref{eq:sc-prob} and the picture of particle emission by a black hole as spontaneous pair creation, was first employed in the study of Hawking radiation  by Srinivasan and Padmanabhan \cite{padmanabhan-99} and, independently, by Parikh and Wilczek \cite{Parikh:1999mf}, and has been considerably refined since then (see Ref.~\cite{Vanzo:2011wq} for a comprehensive review). However, the typical starting point in all these works on black holes is the formula \eqref{eq:sc-prob}, its origin left unexplained. In this and the preceding section, besides obtaining the probability of pair creation by non-singular gravitational collapse, an attempt has been made to provide the link between \eqref{eq:sc-prob} and the understanding of particle emission by black holes as spontaneous pair creation.  

\section{Comparison with the classical result}
The result \eqref{eq:prob-emission} for pair-creation probability, if $\De = 1$, is different from what one would obtain for classical gravitational collapse. This also suggests that the spectrum of radiation from non-singular gravitational collapse of the kind we have considered is not thermal. 

For a concrete comparison, take the case of Vaidya collapse \cite{Ashtekar-Krishnan-2003, Ashtekar:2025-horizons}. As the null fluid collapses, there forms a dynamical horizon, which, at late times, transitions to a non-expanding horizon (NEH), which, in turn, is part of the global event horizon. Following the kind of analysis presented in the preceding section, the pair-creation probability can be shown to assume the following form at late times \cite{Vanzo:2011wq}.
\begin{equation}
    \sigma \sim \exp \left(-\frac{2\pi\om_{\text{NEH}}}{\hbar\kappa_{\text{NEH}}}\right)~.
\end{equation}
The NEH represents an equilibrium configuration, i.e. the surface gravity $\kappa_{\text{NEH}}$ is constant, given by $1/4M_{\text{NEH}}$, where $M_{\text{NEH}}$ is the mass of the NEH as defined, for instance, in \cite{Ashtekar:1998sp}. Spherical symmetry allows us to conclude that $M_{\text{NEH}}=M_{\text{ADM}}$, and also that the spacetime to the future and exterior of the NEH is Schwarzschild, where the Kodama vector field $K^a$ becomes the timelike Killing vector field, and hence $\om_{\text{NEH}} = \om_{\infty}$, the energy of the emitted particle as measured by asymptotic observers. Thus, the pair-creation probability for a mode $\om_\infty$ is constant,
\begin{equation}
    \sigma \sim \exp\left(-8\pi M_{\text{ADM}} ~\om_\infty/\hbar\right) \label{eq:prob-emission-vaidya}
\end{equation}
By invoking the principle of detailed balance \cite[p.~11]{Vanzo:2011wq}, it can further shown to be consistent with a thermal spectrum. To this end, assume that at any time, the NEH and the emitted radiation constitute a system in thermal equilibrium. Let $\la_{N,N-1}\dd t$ denote the conditional probability that the system consists of the black hole plus $N$ quanta given that the system consisted of the black hole plus $N-1$ quanta a time $\dd t$ ago. That is, $\la_{N,N-1}$ is the conditional probability per unit time of the system transitioning from the state ``black hole + $(N-1)$ quanta'' to the state ``black hole + $N$ quanta'' by spontaneous emission. Similarly, $\la_{N-1,N}$ would be the conditional probability per unit time of the system transitioning from the state ``black hole + $N$ quanta'' to the state ``black hole + $(N-1)$ quanta'' by absorption of a quantum. Now consider an ensemble of identical systems each defined as above, and let $\sigma_N$ be the probability that a randomly chosen system out of the ensemble is in the state ``black hole + $N$ quanta''. The principle of detailed balance implies that 
\begin{equation}
    \frac{\la_{N,N-1}}{\la_{N-1,N}} = \frac{\sigma_N}{\sigma_{N-1}} =  \exp\left(-8\pi M_{\text{ADM}} ~\om_\infty/\hbar\right)~,
\end{equation}
where the last equality follows from \eqref{eq:prob-emission-vaidya}. By recursively applying this relation and bearing in mind that $\Sigma_{N=0}^{\infty}\sigma_N = 1$, we can show that the average number of particles emitted in mode $\om_\infty$ is
\begin{equation}
    \braket{N} = \sum_{N=0}^{\infty}N\sigma_N = \frac{1}{e^{8\pi M_{\text{ADM}}\om_{\infty}/\hbar} - 1}~,
\end{equation}
which is a thermal spectrum of bosons at temperature $T=\hbar/8\pi k_{B} M_{\text{ADM}}$. 

The above arguments generally cannot be straightforwardly generalized to non-singular gravitational collapse. While the outer horizon $H_O$ is non-expanding, and can thus be regarded as being in equilibrium, the inner horizon $H_I$, due to the bouncing geometry, necessarily evolves in a very nontrivial manner. Therefore, the trapping horizon may, on the whole, be far from equilibrium. Thus, only in the case ($\De = 0$) of the ingoing antiparticle being absorbed by the black hole before reaching the inner horizon is the probability of pair-creation \eqref{eq:prob-emission} consistent with a thermal spectrum of radiation. For the case $\De = 1$, there is a contribution to the probability from the inner horizon, which being far from equilibrium in general, makes the emitted radiation non-thermal as well. 
\begin{comment}
    It might be possible to determine the precise deviation from thermality if there are regions of the inner horizon where the spacetime geometry is not changing very fast. A look at some explicit examples of non-singular collapse profiles \cite{Husain:2021ojz, Husain:2022gwp, Husain2024} reveals that if the mass $M_{\text{ADM}}$ of the black hole is much larger than the Planck mass, the radius of the inner horizon evolves on the order of $\sim t$, except when it meets the outer horizon. Thus, at intermediary times, one may regard the inner horizon as approximately in equilibrium. In that case, one may invoke a local version of detailed balance and arrive at
\begin{equation}
    \braket{N} = \frac{1}{\exp\left(\frac{2\pi\om_{H_I}}{\kappa_{H_I}} + \frac{2\pi\om_{H_O}}{\kappa_{H_O}} \right) - 1}~.
\end{equation}
Once again, we may write $\kappa_{H_O} = 1/4M_{\text{ADM}}$ and also $\om_{H_O} = \om_{\infty}$ since the geometry outside the outer horizon is static. Thus,
\begin{equation}
    \braket{N} = \frac{1}{e^{8\pi M_{\text{ADM}}\om_{\infty}} e^{2\pi\om_{H_I}/\kappa_{H_I}} - 1}~.
\end{equation}
\end{comment}

Under the assumption of an inner horizon in approximate equilibrium, we may set $\d{M}_{H_I} \approx 0$, whence $\kappa_{H_I} \approx -M^\p_{H_I}/2M_{H_I}$ (cf. \eqref{eq:sg-pg}). Thus, under our assumptions, the deviation from a thermal spectrum depends on how spatially inhomogeneous the inner horizon is. For very small inhomogeneities ($M^\p_{H_I}\approx 0$), the spectrum of emitted radiation is almost thermal. However, this would be true of extremely fine-tuned mass profiles. In fact, a look at the complicated form of the mass function in Ref.~\cite{Hergott:2022hjm} specially engineered to give a smooth bouncing profile suggests that it might not even be possible to construct a nearly homogeneous mass profile that yields the requisite bouncing geometry. In any case, it is certain that no mass profile can tranquilize the highly non-equilibrium behavior of the mass function in the vicinity of where the inner and outer parts of the trapping horizon meet. Therefore, in general, the non-singular collapse emits non-thermal radiation.  

One may expect such non-thermality from some considerations concerning the (in)famous information-loss problem \cite{hawking-information-loss}. If one thinks \cite{Wald:1975kc} of the problem as arising from the fact that, at late times, the partner mode with which every mode of energy incoming at future null infinity is correlated necessarily falls into a classical black hole, and is hence banished into oblivion, then non-singular gravitational collapse involves no loss of information. The resolution of the singularity and the transient nature of the trapping horizon (Fig.~\ref{fig:non-sing-collapse-1}) entail that whatever goes in, also comes out. Thus, all partner modes eventually make it to future null infinity, and correlations between them can, in principle, be measured. Perhaps this explains why the spectrum of radiation is not thermal, for a non-thermal spectrum would come from a non-thermal density matrix for the radiation. A detailed exploration of this matter would be worthwhile. It would involve going beyond a semiclassical derivation of the pair-creation probability and is left for a future work. 

\section{Conclusions and discussion}
Several further questions suggest themselves in view of the preceding analysis. We highlight some of them and offer some tentative remarks concerning each.

First, the tunneling path in Fig.~\ref{fig:pair-creation} is very special. In particular, one assumes the created particle and antiparticle to traverse the spacetime as if they were free particles. This is obviously a simplification. In general, one would expect the gravitational potential in which the particles find themselves to induce possibly multiple scatterings until there is an outgoing particle that eventually escapes to future null infinity. In the context of the path-integral of a relativistic particle, which was the point of reference for our calculations, such scattering events are processes of higher order in $\hbar$ \cite{feynman-49-positrons} and would necessitate going beyond the semiclassical approximation to the propagator \eqref{eq:prop-param-approx}. In addition to these quantum mechanical corrections, there would also be a classical correction to the path of emitted particles/antiparticles coming from reflection off the potential barrier surrounding the black hole. This occurs for particles with high angular momentum and would thus involve dropping the assumption of only radial geodesics. Both of these generalizations are left for a future work. 

Second, given the potential contribution from the inner horizon in the pair-creation probability \eqref{eq:prob-emission}, it is natural to ask whether such a contribution also exists in other types spacetimes that possess an inner horizon. Examples of spacetimes with inner horizons include Hayward \cite{Hayward:2005gi}, Bardeen \cite{bardeen-68}, Dymnikova \cite{Dymnikova:1992ux}, Reissner-Nordstrom, Kerr, and many examples of quantum-corrected Schwarzschild black holes (e.g., \cite{Bonanno:2000ep, Kelly:2020uwj}). The calculation performed in Sec.~\ref{sec:rad-tunneling-methods} can, in principle, be performed for all these spacetimes, resulting in a similar contribution from the inner horizon. Indeed, there exist explicit calculations of radiation from an inner horizon of renormalization-group improved Schwarzschild black holes \cite{Torres:2013cda, Torres:2013cya}, based on tunneling methods, and also from the inner horizon of a Reissner-Nordstrom or Kerr-Newman black hole, based on solving the Klein-Gordon equation near the inner horizon \cite{Peltola:2005xu, WU2005251}. One would expect similar results for the other cases. Thus, it is not altogether surprising that we find an inner-horizon contribution in our calculations. However, it is worth emphasizing that there is a difference between the physical origin of our contribution \eqref{eq:prob-emission} and that in Refs.~\cite{Peltola:2005xu, WU2005251,Torres:2013cda, Torres:2013cya}. In the latter case, the radiation originates from either analytically continuing the solutions to the Klein-Gordon equation across the inner horizon or from a tunneling mechanism near the inner horizon. As such, this means that there is pair-creation near the inner horizon separate from that near the outer horizon. In contrast, the origin of radiation in our calculation is the intersection of the pair-creation path with the inner horizon. There is only one pair-creation event considered in our calculation, namely that which occurs near the outer horizon; it is only that the probability of this event's occurring depends on the fate of the created antiparticle, which may cross the inner horizon, as we saw above. 

Third, in certain LQG-inspired non-singular collapse scenarios based on a dust-gravity system, the shells of matter inevitably collide one another after the matter bounces \cite{Fazzini:2023ova}. The mass function $M(t,r)$ becomes discontinuous at such points since $\lim_{r\to S^+}M > \lim_{r\to S^-}$, where $r=S(t)$ denotes the location of the shell-crossing. Dynamics beyond a shell crossing have to be found out by integrating the equations of motion. This determines how the discontinuity in the mass function at the shell crossing evolves in time -- such traveling discontinuities are called shock waves. There is, in general, an infinity of ways of integrating the equations of motion to find the evolution of the shock wave \cite{Fazzini:2025hsf}. This is encapsulated in the so-called Rankine-Hugoniot (RH) jump condition for the speed of the shock wave, an instance of which we saw in Section~\ref{sec:bh-dynamics}. In our description of the pair-creation process, we have ignored the possibility of shell crossings. It is worth inquiring about the implications of dropping this assumption. We offer some remarks to that end. 

Notice, first of all, that the ingoing antiparticles which are created sufficiently long after the formation of the trapping region will be absorbed by the matter in the rebound phase. Under our assumption that the size of the trapping region $r_{H_O} \gg \sqrt{\hbar}$, there will be many such absorbed antiparticles. Since, furthermore, shell crossings occur for generic initial data a Planck time after the bounce \cite{Fazzini:2023ova}, these antiparticles must be absorbed after the first instance of shell crossing. Finally, because the entire matter profile soon compresses into a thin shell of infinite energy density \cite{Husain:2022gwp}, the ingoing antiparticles are absorbed by the shock wave that forms after the bounce. Now, since an ingoing particle approaches the shock wave from the outside, the result of the absorption process must be a reduction in the value that the mass function function tends to from the right, i.e., $M^+$. Consequently, the jump in the mass function across the shock wave becomes smaller and this also changes the dynamics of the shock wave via the Rankine-Hugoniot condition. It is tempting to speculate that given enough number of ingoing quanta, the discontinuity should be entirely removed, smoothening out the mass function at the shock and thus dispensing with the shock wave altogether. This opens up the possibility of developing quantum-corrected models of non-singular gravitational collapse in a single asymptotic region without shell crossings.   

Finally, it is worth re-emphasizing that the model of radiation from quantum-corrected collapse that has been presented above cannot be trusted in all the regions of spacetime. We do not have access to the precise details of the geometry of spacetime in a region of Planckian dimensions around the bounce. The effective field-theoretic studies on collapse may robustly establish that there is a bounce when matter reaches Planckian densities, but there is no guarantee that one can even describe gravity in terms of a smooth spacetime manifold at such scales. Therefore, all the calculations and conclusions in this paper apply to regions away from the quantum region where the bounce occurs. In particular, the probability of emission \eqref{eq:prob-emission} applies only to antiparticles which cross the inner horizon sufficiently long after the bounce (see Fig.~\ref{fig:pair-creation}). Keeping this in mind, in eqn.\eqref{eq:prob-emission}, $\kappa_{H_I}$ and $\om_{H_{_I}}$ are to be evaluated at times well after the bounce. For the quantum-corrected Oppenheimer-Snyder model in the framework of effective LQG, we saw in Section~\ref{sec:bh-dynamics} that at such times, $\kappa_{H_I}$ can be approximated by eqn.~\eqref{eq:sg-os-after-bounce}. Plugging this into \eqref{eq:prob-emission}, we see that the the inner-horizon correction to the conventional Hawking result looks like
\begin{equation}
    \frac{2\pi\om_{H_I}}{\hbar(t-t_0)}~.
\end{equation}
This suggests that this correction is of the same order in $\hbar$ as the conventional term coming from the outer horizon. Therefore, to the extent that the semiclassical tunneling calculation can be trusted, the extra contribution from the inner horizon cannot be ignored as a higher-order effect which is inconsequential at the scales at which Hawking radiation is significant. Nevertheless, two shortcomings of the result derived in this paper must be clearly pointed. First, as we emphasized at the end of Section~\ref{sec:rad-tunneling-methods}, the assumptions involved in our effective framework do not permit an analysis of spontaneous emission to the future of the outer horizon. Such emission might well have higher order corrections in its spectrum. Second, the result that we have derived can be fully trusted only if it can be reproduced from a more fundamental model of radiation from quantum-corrected collapse. That is, rather than having to put a scalar field on a putative model of collapse which involves quantum corrections to the geometry, such as a bounce instead of a central singularity, one should study a model that involves from the get-go any quantum corrections to both a matter-filled geometry and radiation from that matter. One avenue of developing such a model would be to add a scalar field to the dust-gravity system in the framework of effective LQG \cite{Husain:2022gwp} before taking the semiclassical limit. At the effective level, this would result in coupled partial differential equations for the degrees of freedom of the scalar field and gravity. Solutions to these equations should incorporate corrections to both the geometry and the spectrum of the radiation due to the scalar field.

\section*{Acknowledgments}
This work owes its origin to numerous discussions with Viqar Husain. I am grateful to him, and also to Augusto Cabrera, Francesco Fazzini, Irfan Javed and Edward Wilson-Ewing for valuable feedback and discussions. This work was supported by the Natural Science and Engineering Research Council of Canada.

\nocite{*}

\bibliographystyle{unsrt}
\bibliography{biblio}

@article{hayward-bh-dynamics,
  title = {General laws of black-hole dynamics},
  author = {Hayward, Sean A.},
  journal = {Phys. Rev. D},
  volume = {49},
  issue = {12},
  pages = {6467--6474},
  numpages = {0},
  year = {1994},
  month = {Jun},
  publisher = {American Physical Society},
  doi = {10.1103/PhysRevD.49.6467},
  url = {https://link.aps.org/doi/10.1103/PhysRevD.49.6467}
}

@article{hayward-bh-spherical,
    author = "Hayward, Sean A.",
    title = "{Unified first law of black hole dynamics and relativistic thermodynamics}",
    eprint = "gr-qc/9710089",
    archivePrefix = "arXiv",
    doi = "10.1088/0264-9381/15/10/017",
    journal = "Class. Quant. Grav.",
    volume = "15",
    pages = "3147--3162",
    year = "1998"
}

@article{Vanzo:2011wq,
    author = "Vanzo, L. and Acquaviva, G. and Di Criscienzo, R.",
    title = "{Tunnelling Methods and Hawking's radiation: achievements and prospects}",
    eprint = "1106.4153",
    archivePrefix = "arXiv",
    primaryClass = "gr-qc",
    doi = "10.1088/0264-9381/28/18/183001",
    journal = "Class. Quant. Grav.",
    volume = "28",
    pages = "183001",
    year = "2011"
}

@article{Hergott:2022hjm,
    author = "Hergott, Samantha and Husain, Viqar and Rastgoo, Saeed",
    title = "{Model metrics for quantum black hole evolution: Gravitational collapse, singularity resolution, and transient horizons}",
    eprint = "2206.06425",
    archivePrefix = "arXiv",
    primaryClass = "gr-qc",
    doi = "10.1103/PhysRevD.106.046012",
    journal = "Phys. Rev. D",
    volume = "106",
    number = "4",
    pages = "046012",
    year = "2022"
}

@article{Singh:2011fb,
    author = "Singh, Suprit and Padmanabhan, T.",
    title = "{Complex Effective Path: A Semi-Classical Probe of Quantum Effects}",
    eprint = "1112.6279",
    archivePrefix = "arXiv",
    primaryClass = "hep-th",
    doi = "10.1103/PhysRevD.85.025011",
    journal = "Phys. Rev. D",
    volume = "85",
    pages = "025011",
    year = "2012"
}

@article{Feynman:1950ir,
    author = "Feynman, R. P.",
    editor = "Brown, L. M.",
    title = "{Mathematical formulation of the quantum theory of electromagnetic interaction}",
    doi = "10.1103/PhysRev.80.440",
    journal = "Phys. Rev.",
    volume = "80",
    pages = "440--457",
    year = "1950",
}

@article{Hartle:1976tp,
    author = "Hartle, J. B. and Hawking, S. W.",
    title = "{Path Integral Derivation of Black Hole Radiance}",
    doi = "10.1103/PhysRevD.13.2188",
    journal = "Phys. Rev. D",
    volume = "13",
    pages = "2188--2203",
    year = "1976"
}

@inbook{Landau:1991wop,
    author = "Landau, Lev Davidovich and Lifshits, E. M.",
    title = "{Quantum Mechanics}: {Non-Relativistic Theory}",
    edition = {3rd},
    isbn = "978-0-7506-3539-4",
    publisher = "Pergamon Press",
    address = "Oxford",
    series = "Course of Theoretical Physics",
    volume = "3",
    year = "1991",
    chapter = {VII}
}

@article{padmanabhan-99,
  title = {Particle production and complex path analysis},
  author = {Srinivasan, K. and Padmanabhan, T.},
  journal = {Phys. Rev. D},
  volume = {60},
  issue = {2},
  pages = {024007},
  numpages = {20},
  year = {1999},
  month = {Jun},
  publisher = {American Physical Society},
  doi = {10.1103/PhysRevD.60.024007},
  url = {https://link.aps.org/doi/10.1103/PhysRevD.60.024007}
}

@article{Husain:2022gwp,
    author = "Husain, Viqar and Kelly, Jarod George and Santacruz, Robert and Wilson-Ewing, Edward",
    title = "{Fate of quantum black holes}",
    eprint = "2203.04238",
    archivePrefix = "arXiv",
    primaryClass = "gr-qc",
    doi = "10.1103/PhysRevD.106.024014",
    journal = "Phys. Rev. D",
    volume = "106",
    number = "2",
    pages = "024014",
    year = "2022"
}

@article{Parikh:1999mf,
    author = "Parikh, Maulik K. and Wilczek, Frank",
    title = "{Hawking radiation as tunneling}",
    eprint = "hep-th/9907001",
    archivePrefix = "arXiv",
    reportNumber = "PUPT-1775, SPIN-1998-12, IASSNS-HEP-98-22",
    doi = "10.1103/PhysRevLett.85.5042",
    journal = "Phys. Rev. Lett.",
    volume = "85",
    pages = "5042--5045",
    year = "2000"
}

@article{Hawking:1975vcx,
    author = "Hawking, S. W.",
    editor = "Gibbons, G. W. and Hawking, S. W.",
    title = "{Particle Creation by Black Holes}",
    doi = "10.1007/BF02345020",
    journal = "Commun. Math. Phys.",
    volume = "43",
    pages = "199--220",
    year = "1975",
    note = "[Erratum: Commun.Math.Phys. 46, 206 (1976)]"
}

@article{Ashtekar:2025ptw,
    author = "Ashtekar, Abhay",
    title = "{Black hole evaporation in loop quantum gravity}",
    eprint = "2502.04252",
    archivePrefix = "arXiv",
    primaryClass = "gr-qc",
    doi = "10.1007/s10714-025-03380-7",
    journal = "Gen. Rel. Grav.",
    volume = "57",
    number = "2",
    pages = "48",
    year = "2025"
}

@article{Parker:1968mv,
    author = "Parker, L.",
    title = "{Particle creation in expanding universes}",
    doi = "10.1103/PhysRevLett.21.562",
    journal = "Phys. Rev. Lett.",
    volume = "21",
    pages = "562--564",
    year = "1968"
}

@article{Imamura:1960tzx,
    author = "Imamura, Tsutomu",
    title = "{Quantized Meson Field in a Classical Gravitational Field}",
    doi = "10.1103/PhysRev.118.1430",
    journal = "Phys. Rev.",
    volume = "118",
    number = "5",
    pages = "1430--1434",
    year = "1960"
}

@article{Hawking:1994ss,
    author = "Hawking, S. W.",
    title = "{Nature of space and time}",
    eprint = "hep-th/9409195",
    archivePrefix = "arXiv",
    month = "9",
    year = "1994"
}

@article{Fredenhagen:1989kr,
    author = "Fredenhagen, Klaus and Haag, Rudolf",
    title = "{On the Derivation of Hawking Radiation Associated With the Formation of a Black Hole}",
    reportNumber = "DESY-89-045",
    doi = "10.1007/BF02096757",
    journal = "Commun. Math. Phys.",
    volume = "127",
    pages = "273",
    year = "1990"
}

@article{schwinger-1951,
  title = {On Gauge Invariance and Vacuum Polarization},
  author = {Schwinger, Julian},
  journal = {Phys. Rev.},
  volume = {82},
  issue = {5},
  pages = {664--679},
  numpages = {0},
  year = {1951},
  month = {Jun},
  publisher = {American Physical Society},
  doi = {10.1103/PhysRev.82.664},
  url = {https://link.aps.org/doi/10.1103/PhysRev.82.664}
}

@article{Heisenberg:1936nmg,
    author = "Heisenberg, W. and Euler, H.",
    title = "{Consequences of Dirac's theory of positrons}",
    eprint = "physics/0605038",
    archivePrefix = "arXiv",
    doi = "10.1007/BF01343663",
    journal = "Z. Phys.",
    volume = "98",
    number = "11-12",
    pages = "714--732",
    year = "1936"
}

@article{Sauter-1931,
    author = {Sauter, Fritz},
    title = {Über das Verhalten eines Elektrons im homogenen elektrischen Feld nach der relativistischen Theorie Diracs},
    journal = {Zeitschrift für Physik},
    volume = {69},
    number = {11},
    pages = {742-764},
    year = {1931},
    doi = {10.1007/BF01339461},
    url = {https://doi.org/10.1007/BF01339461}
}

@article{choptuik-1993,
  title = {Universality and scaling in gravitational collapse of a massless scalar field},
  author = {Choptuik, Matthew W.},
  journal = {Phys. Rev. Lett.},
  volume = {70},
  issue = {1},
  pages = {9--12},
  numpages = {0},
  year = {1993},
  month = {Jan},
  publisher = {American Physical Society},
  doi = {10.1103/PhysRevLett.70.9},
  url = {https://link.aps.org/doi/10.1103/PhysRevLett.70.9}
}

@inproceedings{Wald:1997wa,
    author = "Wald, Robert M.",
    title = "{Gravitational collapse and cosmic censorship}",
    eprint = "gr-qc/9710068",
    archivePrefix = "arXiv",
    reportNumber = "EFI-97-43",
    doi = "10.1007/978-94-017-0934-7_5",
    pages = "69--85",
    month = "10",
    year = "1997"
}

@article{Christodoulou-99,
 ISSN = {0003486X},
 URL = {http://www.jstor.org/stable/121023},
 author = {Demetrios Christodoulou},
 journal = {Annals of Mathematics},
 number = {1},
 pages = {183--217},
 publisher = {Annals of Mathematics},
 title = {The Instability of Naked Singularities in the Gravitational Collapse of a Scalar Field},
 urldate = {2025-07-17},
 volume = {149},
 year = {1999}
}

@article{Witten:2024upt,
    author = "Witten, Edward",
    title = "{Introduction to black hole thermodynamics}",
    eprint = "2412.16795",
    archivePrefix = "arXiv",
    primaryClass = "hep-th",
    doi = "10.1140/epjp/s13360-025-06288-y",
    journal = "Eur. Phys. J. Plus",
    volume = "140",
    number = "5",
    pages = "430",
    year = "2025"
}

@article{Padmanabhan:2009vy,
    author = "Padmanabhan, T.",
    title = "{Thermodynamical Aspects of Gravity: New insights}",
    eprint = "0911.5004",
    archivePrefix = "arXiv",
    primaryClass = "gr-qc",
    doi = "10.1088/0034-4885/73/4/046901",
    journal = "Rept. Prog. Phys.",
    volume = "73",
    pages = "046901",
    year = "2010"
}

@article{Wilson-Ewing:2016yan,
    author = "Wilson-Ewing, Edward",
    title = "{Testing loop quantum cosmology}",
    eprint = "1612.04551",
    archivePrefix = "arXiv",
    primaryClass = "gr-qc",
    doi = "10.1016/j.crhy.2017.02.004",
    journal = "Comptes Rendus Physique",
    volume = "18",
    pages = "207--225",
    year = "2017"
}

@article{Ashtekar:2011ni,
    author = "Ashtekar, Abhay and Singh, Parampreet",
    title = "{Loop Quantum Cosmology: A Status Report}",
    eprint = "1108.0893",
    archivePrefix = "arXiv",
    primaryClass = "gr-qc",
    doi = "10.1088/0264-9381/28/21/213001",
    journal = "Class. Quant. Grav.",
    volume = "28",
    pages = "213001",
    year = "2011"
}

@article{Husain:2021ojz,
    author = "Husain, Viqar and Kelly, Jarod George and Santacruz, Robert and Wilson-Ewing, Edward",
    title = "{Quantum Gravity of Dust Collapse: Shock Waves from Black Holes}",
    eprint = "2109.08667",
    archivePrefix = "arXiv",
    primaryClass = "gr-qc",
    doi = "10.1103/PhysRevLett.128.121301",
    journal = "Phys. Rev. Lett.",
    volume = "128",
    number = "12",
    pages = "121301",
    year = "2022"
}

@Inbook{Husain2024,
author="Husain, Viqar",
editor="Malafarina, Daniele
and Joshi, Pankaj S.",
title="Quantum Black Holes: A Survey",
bookTitle="New Frontiers in Gravitational Collapse and Spacetime Singularities",
year="2024",
publisher="Springer Nature Singapore",
address="Singapore",
pages="101--124",
doi="10.1007/978-981-97-1172-7_4",
url="https://doi.org/10.1007/978-981-97-1172-7_4"
}

@article{Cipriani:2024nhx,
    author = "Cipriani, Lorenzo and Fazzini, Francesco and Wilson-Ewing, Edward",
    title = "{Gravitational collapse in effective loop quantum gravity: Beyond marginally bound configurations}",
    eprint = "2404.04192",
    archivePrefix = "arXiv",
    primaryClass = "gr-qc",
    doi = "10.1103/PhysRevD.110.066004",
    journal = "Phys. Rev. D",
    volume = "110",
    number = "6",
    pages = "066004",
    year = "2024"
}

@article{Bojowald:2024ium,
    author = "Bojowald, Martin and Duque, Erick I. and Hartmann, Dennis",
    title = "{Covariant Lema{\^\i}tre-Tolman-Bondi collapse in models of loop quantum gravity}",
    eprint = "2412.18054",
    archivePrefix = "arXiv",
    primaryClass = "gr-qc",
    doi = "10.1103/PhysRevD.111.064002",
    journal = "Phys. Rev. D",
    volume = "111",
    number = "6",
    pages = "064002",
    year = "2025"
}

@article{Ashtekar-Krishnan-2003,
  title = {Dynamical horizons and their properties},
  author = {Ashtekar, Abhay and Krishnan, Badri},
  journal = {Phys. Rev. D},
  volume = {68},
  issue = {10},
  pages = {104030},
  numpages = {25},
  year = {2003},
  month = {Nov},
  publisher = {American Physical Society},
  doi = {10.1103/PhysRevD.68.104030},
  url = {https://link.aps.org/doi/10.1103/PhysRevD.68.104030}
}

@article{Ashtekar:2025-horizons,
    author = "Ashtekar, Abhay and Krishnan, Badri",
    title = "{Quasi-Local Black Hole Horizons: Recent Advances}",
    eprint = "2502.11825",
    archivePrefix = "arXiv",
    primaryClass = "gr-qc",
    month = "2",
    year = "2025"
}

@article{Bondi:1947fta,
    author = "Bondi, H.",
    title = "{Spherically symmetrical models in general relativity}",
    doi = "10.1093/mnras/107.5-6.410",
    journal = "Mon. Not. Roy. Astron. Soc.",
    volume = "107",
    pages = "410--425",
    year = "1947"
}

@article{Lemaitre:1931zza,
    author = "Lemaitre, Georges",
    title = "{A Homogeneous Universe of Constant Mass and Increasing Radius accounting for the Radial Velocity of Extra-galactic Nebul{\ae}}",
    doi = "10.1093/mnras/91.5.483",
    journal = "Mon. Not. Roy. Astron. Soc.",
    volume = "91",
    number = "5",
    pages = "483--490",
    year = "1931"
}

@article{Tolman:1934za,
    author = "Tolman, Richard C.",
    title = "{Effect of imhomogeneity on cosmological models}",
    doi = "10.1073/pnas.20.3.169",
    journal = "Proc. Nat. Acad. Sci.",
    volume = "20",
    pages = "169--176",
    year = "1934"
}

@article{Kodama:1979vn,
    author = "Kodama, Hideo",
    title = "{Conserved Energy Flux for the Spherically Symmetric System and the Back Reaction Problem in the Black Hole Evaporation}",
    reportNumber = "KUNS-506",
    doi = "10.1143/PTP.63.1217",
    journal = "Prog. Theor. Phys.",
    volume = "63",
    pages = "1217",
    year = "1980"
}

@article{stephens-1989,
  author       = {Stephens, C R},
  title        = {The Hawking effect in abelian gauge theories},
  doi          = {10.1016/0003-4916(89)90001-8},
  url          = {https://www.osti.gov/biblio/5350209},
  journal      = {Annals of Physics (New York); (USA)},
  issn         = {ISSN 0003-4916},
  volume       = {193:2},
  place        = {United States},
  year         = {1989},
  month        = {08}}

@article{BROUT1991209,
title = {Thermal properties of pairs produced by an electric field: A tunneling approach},
journal = {Nuclear Physics B},
volume = {353},
number = {1},
pages = {209-236},
year = {1991},
issn = {0550-3213},
doi = {https://doi.org/10.1016/0550-3213(91)90508-U},
url = {https://www.sciencedirect.com/science/article/pii/055032139190508U},
author = {R. Brout and R. Parentani and Ph. Spindel},
}

@article{Damour:1976jd,
    author = "Damour, T. and Ruffini, R.",
    title = "{Black Hole Evaporation in the Klein-Sauter-Heisenberg-Euler Formalism}",
    doi = "10.1103/PhysRevD.14.332",
    journal = "Phys. Rev. D",
    volume = "14",
    pages = "332--334",
    year = "1976"
}

@article{Wondrak:2023zdi,
    author = "Wondrak, Michael F. and van Suijlekom, Walter D. and Falcke, Heino",
    title = "{Gravitational Pair Production and Black Hole Evaporation}",
    eprint = "2305.18521",
    archivePrefix = "arXiv",
    primaryClass = "gr-qc",
    doi = "10.1103/PhysRevLett.130.221502",
    journal = "Phys. Rev. Lett.",
    volume = "130",
    number = "22",
    pages = "221502",
    year = "2023"
}

@article{Ashtekar:1998sp,
    author = "Ashtekar, Abhay and Beetle, Christopher and Fairhurst, Stephen",
    title = "{Isolated horizons: A Generalization of black hole mechanics}",
    eprint = "gr-qc/9812065",
    archivePrefix = "arXiv",
    doi = "10.1088/0264-9381/16/2/027",
    journal = "Class. Quant. Grav.",
    volume = "16",
    pages = "L1--L7",
    year = "1999"
}

@article{hawking-information-loss,
  title = {Breakdown of predictability in gravitational collapse},
  author = {Hawking, S. W.},
  journal = {Phys. Rev. D},
  volume = {14},
  issue = {10},
  pages = {2460--2473},
  numpages = {0},
  year = {1976},
  month = {Nov},
  publisher = {American Physical Society},
  doi = {10.1103/PhysRevD.14.2460},
  url = {https://link.aps.org/doi/10.1103/PhysRevD.14.2460}
}

@article{Wald:1975kc,
    author = "Wald, Robert M.",
    title = "{On Particle Creation by Black Holes}",
    doi = "10.1007/BF01609863",
    journal = "Commun. Math. Phys.",
    volume = "45",
    pages = "9--34",
    year = "1975"
}

@article{Fock:1937dy,
    author = "Fock, V.",
    title = "{Proper time in classical and quantum mechanics}",
    journal = "Phys. Z. Sowjetunion",
    volume = "12",
    pages = "404--425",
    year = "1937"
}

@article{Nambu:1950rs,
    author = "Nambu, Yoichiro",
    title = "{The use of the Proper Time in Quantum Electrodynamics}",
    doi = "10.1143/PTP.5.82",
    journal = "Prog. Theor. Phys.",
    volume = "5",
    pages = "82--94",
    year = "1950"
}

@article{Stueckelberg:1941rg,
    author = "Stueckelberg, E. C. G.",
    title = "{Remarks on the creation of pairs of particles in the theory of relativity}",
    journal = "Helv. Phys. Acta",
    volume = "14",
    pages = "588--594",
    year = "1941"
}

@article{Morette:1951zz,
    author = "Morette, C.",
    title = "{On the definition and approximation of Feynman's path integrals}",
    doi = "10.1103/PhysRev.81.848",
    journal = "Phys. Rev.",
    volume = "81",
    pages = "848--852",
    year = "1951"
}

@article{holstein-swift-1,
    author = {Holstein, Barry R. and Swift, Arthur R.},
    title = {Path integrals and the WKB approximation},
    journal = {American Journal of Physics},
    volume = {50},
    number = {9},
    pages = {829-832},
    year = {1982},
    month = {09},
    issn = {0002-9505},
    doi = {10.1119/1.12750},
    url = {https://doi.org/10.1119/1.12750},
    eprint = {https://pubs.aip.org/aapt/ajp/article-pdf/50/9/829/11552227/829\_1\_online.pdf},
}

@article{holstein-swift-2,
    author = {Holstein, Barry R. and Swift, Arthur R.},
    title = {Barrier penetration via path integrals},
    journal = {American Journal of Physics},
    volume = {50},
    number = {9},
    pages = {833-839},
    year = {1982},
    month = {09},
    issn = {0002-9505},
    doi = {10.1119/1.12751},
    url = {https://doi.org/10.1119/1.12751},
    eprint = {https://pubs.aip.org/aapt/ajp/article-pdf/50/9/833/11552245/833\_1\_online.pdf},
}

@article{DiCriscienzo:2009kun,
    author = "Di Criscienzo, Roberto and Hayward, Sean A. and Nadalini, Mario and Vanzo, Luciano and Zerbini, Sergio",
    title = "{Hamilton-Jacobi tunneling method for dynamical horizons in different coordinate gauges}",
    eprint = "0906.1725",
    archivePrefix = "arXiv",
    primaryClass = "gr-qc",
    doi = "10.1088/0264-9381/27/1/015006",
    journal = "Class. Quant. Grav.",
    volume = "27",
    pages = "015006",
    year = "2010"
}

@article{Hayward:2008jq,
    author = "Hayward, S. A. and Di Criscienzo, R. and Vanzo, L. and Nadalini, M. and Zerbini, S.",
    title = "{Local Hawking temperature for dynamical black holes}",
    eprint = "0806.0014",
    archivePrefix = "arXiv",
    primaryClass = "gr-qc",
    doi = "10.1088/0264-9381/26/6/062001",
    journal = "Class. Quant. Grav.",
    volume = "26",
    pages = "062001",
    year = "2009"
}

@article{Giesel:2023hys,
    author = "Giesel, Kristina and Liu, Hongguang and Singh, Parampreet and Weigl, Stefan Andreas",
    title = "{Generalized analysis of a dust collapse in effective loop quantum gravity: Fate of shocks and covariance}",
    eprint = "2308.10953",
    archivePrefix = "arXiv",
    primaryClass = "gr-qc",
    doi = "10.1103/PhysRevD.110.104016",
    journal = "Phys. Rev. D",
    volume = "110",
    number = "10",
    pages = "104016",
    year = "2024"
}

@article{Fazzini:2023ova,
    author = "Fazzini, Francesco and Husain, Viqar and Wilson-Ewing, Edward",
    title = "{Shell-crossings and shock formation during gravitational collapse in effective loop quantum gravity}",
    eprint = "2312.02032",
    archivePrefix = "arXiv",
    primaryClass = "gr-qc",
    doi = "10.1103/PhysRevD.109.084052",
    journal = "Phys. Rev. D",
    volume = "109",
    number = "8",
    pages = "084052",
    year = "2024"
}

@article{ford-parker-78,
  title = {Creation of particles by singularities in asymptotically flat spacetimes},
  author = {Ford, L. H. and Parker, Leonard},
  journal = {Phys. Rev. D},
  volume = {17},
  issue = {6},
  pages = {1485--1496},
  numpages = {0},
  year = {1978},
  month = {Mar},
  publisher = {American Physical Society},
  doi = {10.1103/PhysRevD.17.1485},
  url = {https://link.aps.org/doi/10.1103/PhysRevD.17.1485}
}

@article{feynman-49-positrons,
  title = {The Theory of Positrons},
  author = {Feynman, R. P.},
  journal = {Phys. Rev.},
  volume = {76},
  issue = {6},
  pages = {749--759},
  numpages = {0},
  year = {1949},
  month = {Sep},
  publisher = {American Physical Society},
  doi = {10.1103/PhysRev.76.749},
  url = {https://link.aps.org/doi/10.1103/PhysRev.76.749}
}

@article{Fazzini:2025hsf,
    author = "Fazzini, Francesco",
    title = "{Non-uniqueness of the shockwave dynamics in effective loop quantum gravity}",
    eprint = "2502.03003",
    archivePrefix = "arXiv",
    primaryClass = "gr-qc",
    month = "2",
    year = "2025"
}

@article{Dymnikova:1992ux,
    author = "Dymnikova, I.",
    title = "{Vacuum nonsingular black hole}",
    doi = "10.1007/BF00760226",
    journal = "Gen. Rel. Grav.",
    volume = "24",
    pages = "235--242",
    year = "1992"
}

@article{Hayward:2005gi,
    author = "Hayward, Sean A.",
    title = "{Formation and evaporation of regular black holes}",
    eprint = "gr-qc/0506126",
    archivePrefix = "arXiv",
    doi = "10.1103/PhysRevLett.96.031103",
    journal = "Phys. Rev. Lett.",
    volume = "96",
    pages = "031103",
    year = "2006"
}

@article{WU2005251,
title = {New formulation of the first law of black hole thermodynamics: a stringy analogy},
journal = {Physics Letters B},
volume = {608},
number = {3},
pages = {251-257},
year = {2005},
issn = {0370-2693},
doi = {https://doi.org/10.1016/j.physletb.2005.01.018},
url = {https://www.sciencedirect.com/science/article/pii/S0370269305000377},
author = {Shuang-Qing Wu},
}

@article{Peltola:2005xu,
    author = "Peltola, Ari and Makela, Jarmo",
    title = {{Radiation of the inner horizon of the Reissner-Nordstr{\"o}m black hole}},
    eprint = "gr-qc/0508095",
    archivePrefix = "arXiv",
    doi = "10.1142/S0218271806008565",
    journal = "Int. J. Mod. Phys. D",
    volume = "15",
    pages = "817--844",
    year = "2006"
}

@article{bardeen-68,
    author = "Bardeen, James",
    title = "{Nonsingular general relativistic gravitational collapse}",
    year = {1968}
}

@article{Saini:2014qpa,
    author = "Saini, Anshul and Stojkovic, Dejan",
    title = "{Nonlocal (but also nonsingular) physics at the last stages of gravitational collapse}",
    eprint = "1401.6182",
    archivePrefix = "arXiv",
    primaryClass = "gr-qc",
    doi = "10.1103/PhysRevD.89.044003",
    journal = "Phys. Rev. D",
    volume = "89",
    number = "4",
    pages = "044003",
    year = "2014"
}

@article{Greenwood:2008ht,
    author = "Greenwood, Eric and Stojkovic, Dejan",
    title = "{Quantum gravitational collapse: Non-singularity and non-locality}",
    eprint = "0802.4087",
    archivePrefix = "arXiv",
    primaryClass = "gr-qc",
    doi = "10.1088/1126-6708/2008/06/042",
    journal = "JHEP",
    volume = "06",
    pages = "042",
    year = "2008"
}

@article{Wang:2009ay,
    author = "Wang, John E. and Greenwood, Eric and Stojkovic, Dejan",
    title = "{Schrodinger formalism, black hole horizons and singularity behavior}",
    eprint = "0906.3250",
    archivePrefix = "arXiv",
    primaryClass = "hep-th",
    doi = "10.1103/PhysRevD.80.124027",
    journal = "Phys. Rev. D",
    volume = "80",
    pages = "124027",
    year = "2009"
}

@article{Bambi:2013gva,
    author = "Bambi, Cosimo and Malafarina, Daniele and Modesto, Leonardo",
    title = "{Terminating black holes in asymptotically free quantum gravity}",
    eprint = "1306.1668",
    archivePrefix = "arXiv",
    primaryClass = "gr-qc",
    doi = "10.1140/epjc/s10052-014-2767-9",
    journal = "Eur. Phys. J. C",
    volume = "74",
    pages = "2767",
    year = "2014"
}

@article{Bambi:2016uda,
    author = "Bambi, Cosimo and Malafarina, Daniele and Modesto, Leonardo",
    title = "{Black supernovae and black holes in non-local gravity}",
    eprint = "1603.09592",
    archivePrefix = "arXiv",
    primaryClass = "gr-qc",
    doi = "10.1007/JHEP04(2016)147",
    journal = "JHEP",
    volume = "04",
    pages = "147",
    year = "2016"
}

@article{Liu:2014kra,
    author = "Liu, Yue and Malafarina, Daniele and Modesto, Leonardo and Bambi, Cosimo",
    title = "{Singularity avoidance in quantum-inspired inhomogeneous dust collapse}",
    eprint = "1405.7249",
    archivePrefix = "arXiv",
    primaryClass = "gr-qc",
    doi = "10.1103/PhysRevD.90.044040",
    journal = "Phys. Rev. D",
    volume = "90",
    number = "4",
    pages = "044040",
    year = "2014"
}

@article{Torres:2014pea,
    author = "Torres, R. and Fayos, F.",
    title = "{Singularity free gravitational collapse in an effective dynamical quantum spacetime}",
    eprint = "1405.7922",
    archivePrefix = "arXiv",
    primaryClass = "gr-qc",
    doi = "10.1016/j.physletb.2014.04.038",
    journal = "Phys. Lett. B",
    volume = "733",
    pages = "169--175",
    year = "2014"
}

@article{Torres:2013cda,
    author = "Torres, R.",
    title = "{On the interior of (Quantum) Black Holes}",
    eprint = "1309.1083",
    archivePrefix = "arXiv",
    primaryClass = "gr-qc",
    doi = "10.1016/j.physletb.2013.06.031",
    journal = "Phys. Lett. B",
    volume = "724",
    pages = "338--345",
    year = "2013"
}

@article{Torres:2013cya,
    author = "Torres, R. and Fayos, F. and Lorente-Espin, O.",
    title = "{Evaporation of (quantum) black holes and energy conservation}",
    eprint = "1308.4318",
    archivePrefix = "arXiv",
    primaryClass = "gr-qc",
    doi = "10.1016/j.physletb.2013.01.061",
    journal = "Phys. Lett. B",
    volume = "720",
    pages = "198--204",
    year = "2013"
}

@article{Haggard:2014rza,
    author = "Haggard, Hal M. and Rovelli, Carlo",
    title = "{Quantum-gravity effects outside the horizon spark black to white hole tunneling}",
    eprint = "1407.0989",
    archivePrefix = "arXiv",
    primaryClass = "gr-qc",
    doi = "10.1103/PhysRevD.92.104020",
    journal = "Phys. Rev. D",
    volume = "92",
    number = "10",
    pages = "104020",
    year = "2015"
}

@article{Han:2023wxg,
    author = "Han, Muxin and Rovelli, Carlo and Soltani, Farshid",
    title = "{Geometry of the black-to-white hole transition within a single asymptotic region}",
    eprint = "2302.03872",
    archivePrefix = "arXiv",
    primaryClass = "gr-qc",
    doi = "10.1103/PhysRevD.107.064011",
    journal = "Phys. Rev. D",
    volume = "107",
    number = "6",
    pages = "064011",
    year = "2023"
}

@inbook{arfken-2013,
    author = {George B. Arfken and Hans J. Weber and Frank E. Harris},
    title = {Mathematical Methods for Physicists},
    edition = {7th},
    publisher = {Academic Press},
    year = {2013},
    chapter = {11},
    isbn = {978-0-12-384654-9},
    address = {Boston},
    pages = {469-550},
}

@article{Kelly:2020uwj,
    author = "Kelly, Jarod George and Santacruz, Robert and Wilson-Ewing, Edward",
    title = "{Effective loop quantum gravity framework for vacuum spherically symmetric spacetimes}",
    eprint = "2006.09302",
    archivePrefix = "arXiv",
    primaryClass = "gr-qc",
    doi = "10.1103/PhysRevD.102.106024",
    journal = "Phys. Rev. D",
    volume = "102",
    number = "10",
    pages = "106024",
    year = "2020"
}

@article{Bonanno:2000ep,
    author = "Bonanno, Alfio and Reuter, Martin",
    title = "{Renormalization group improved black hole space-times}",
    eprint = "hep-th/0002196",
    archivePrefix = "arXiv",
    reportNumber = "INFN-CT-03-00, MZ-TH-00-04",
    doi = "10.1103/PhysRevD.62.043008",
    journal = "Phys. Rev. D",
    volume = "62",
    pages = "043008",
    year = "2000"
}

@article{Ashtekar:2004eh,
    author = "Ashtekar, Abhay and Lewandowski, Jerzy",
    title = "{Background independent quantum gravity: A Status report}",
    eprint = "gr-qc/0404018",
    archivePrefix = "arXiv",
    doi = "10.1088/0264-9381/21/15/R01",
    journal = "Class. Quant. Grav.",
    volume = "21",
    pages = "R53",
    year = "2004"
}

@book{Ashtekar:2017yom,
    editor = "Ashtekar, Abhay and Pullin, Jorge",
    title = "{Loop Quantum Gravity}: {The First 30 Years}",
    doi = "10.1142/10445",
    isbn = "978-981-320-992-3, 978-981-322-001-0, 978-981-320-993-0",
    publisher = "World Scientific",
    series = "100 Years of General Relativity",
    volume = "4",
    year = "2017"
}

@article{Thiemann:2000bw,
    author = "Thiemann, Thomas",
    title = "{Gauge field theory coherent states (GCS): 1. General properties}",
    eprint = "hep-th/0005233",
    archivePrefix = "arXiv",
    reportNumber = "AEI-2000-027",
    doi = "10.1088/0264-9381/18/11/304",
    journal = "Class. Quant. Grav.",
    volume = "18",
    pages = "2025--2064",
    year = "2001"
}

@article{Husain:2006uh,
    author = "Husain, Viqar and Winkler, Oliver",
    title = "{Semiclassical states for quantum cosmology}",
    eprint = "gr-qc/0607097",
    archivePrefix = "arXiv",
    doi = "10.1103/PhysRevD.75.024014",
    journal = "Phys. Rev. D",
    volume = "75",
    pages = "024014",
    year = "2007"
}

@article{Bojowald:2012xy,
    author = "Bojowald, Martin",
    title = "{Quantum Cosmology: Effective Theory}",
    eprint = "1209.3403",
    archivePrefix = "arXiv",
    primaryClass = "gr-qc",
    doi = "10.1088/0264-9381/29/21/213001",
    journal = "Class. Quant. Grav.",
    volume = "29",
    pages = "213001",
    year = "2012"
}

@article{Zhang:2024khj,
    author = "Zhang, Cong and Lewandowski, Jerzy and Ma, Yongge and Yang, Jinsong",
    title = "{Black holes and covariance in effective quantum gravity}",
    eprint = "2407.10168",
    archivePrefix = "arXiv",
    primaryClass = "gr-qc",
    doi = "10.1103/PhysRevD.111.L081504",
    journal = "Phys. Rev. D",
    volume = "111",
    number = "8",
    pages = "L081504",
    year = "2025"
}

@article{Giesel:2023tsj,
    author = "Giesel, Kristina and Liu, Hongguang and Rullit, Eric and Singh, Parampreet and Weigl, Stefan Andreas",
    title = "{Embedding generalized Lema{\^\i}tre-Tolman-Bondi models in polymerized spherically symmetric spacetimes}",
    eprint = "2308.10949",
    archivePrefix = "arXiv",
    primaryClass = "gr-qc",
    doi = "10.1103/PhysRevD.110.104017",
    journal = "Phys. Rev. D",
    volume = "110",
    number = "10",
    pages = "104017",
    year = "2024"
}

@article{Alonso-Bardaji:2025hda,
    author = "Alonso-Bardaji, Asier and Brizuela, David",
    title = "{Dynamical theory for spherical black holes in modified gravity}",
    eprint = "2507.19380",
    archivePrefix = "arXiv",
    primaryClass = "gr-qc",
    doi = "10.1103/ttfy-yjdh",
    journal = "Phys. Rev. D",
    volume = "112",
    number = "10",
    pages = "104036",
    year = "2025"
}

@book{leveque,
  author = {LeVeque, Randall J.},
  isbn = {978-3-7643-2723-1},
  pages = {1-214},
  publisher = {Birkhäuser},
  series = {Lectures in mathematics},
  title = {Numerical methods for conservation laws (2. ed.).},
  year = 1992
}

@article{Callan:1992rs,
    author = "Callan, Jr., Curtis G. and Giddings, Steven B. and Harvey, Jeffrey A. and Strominger, Andrew",
    title = "{Evanescent black holes}",
    eprint = "hep-th/9111056",
    archivePrefix = "arXiv",
    reportNumber = "UCSB-TH-91-54, EFI-91-67, PUPT-1294",
    doi = "10.1103/PhysRevD.45.R1005",
    journal = "Phys. Rev. D",
    volume = "45",
    number = "4",
    pages = "R1005",
    year = "1992"
}

@article{Hossenfelder:2009xq,
    author = "Hossenfelder, Sabine and Smolin, Lee",
    title = "{Conservative solutions to the black hole information problem}",
    eprint = "0901.3156",
    archivePrefix = "arXiv",
    primaryClass = "gr-qc",
    doi = "10.1103/PhysRevD.81.064009",
    journal = "Phys. Rev. D",
    volume = "81",
    pages = "064009",
    year = "2010"
}

@article{Boasso:2024ryt,
    author = "Boasso, Andr{\'e}s and Franchino-Vi{\~n}as, Sebasti{\'a}n and Mazzitelli, Francisco D.",
    title = "{Nonlocal effective action and particle creation in D dimensions}",
    eprint = "2412.03340",
    archivePrefix = "arXiv",
    primaryClass = "hep-th",
    doi = "10.1103/PhysRevD.111.085023",
    journal = "Phys. Rev. D",
    volume = "111",
    number = "8",
    pages = "085023",
    year = "2025"
}

\end{document}